\documentclass[a4paper,11pt]{article}

\usepackage{amsmath,amsthm,amsbsy,amssymb,amsfonts,graphicx,mathrsfs,bm,cite,color}
\usepackage[unicode]{hyperref}

\makeatletter
\catcode`\@=11
\@addtoreset{equation}{section}

\makeatother

\usepackage[utf8]{inputenc}

\renewcommand{\thefootnote}{\arabic{footnote}}

\newcommand{\Exp}[1]{\operatorname{e}^{#1}}
\newcommand{\abs}[1]{\lvert {#1} \rvert}
\newcommand{\rmd}{{\mathrm{d}}}

\newcommand{\Lie}{\pounds}

\newcommand{\cD}{\mathcal D}
\newcommand{\cF}{\mathcal F}
\newcommand{\cH}{\mathcal H}

\newcommand{\cM}{\mathcal M}

\newcommand{\SL}{\text{SL}}
\newcommand{\SU}{\text{SU}}
\newcommand{\OO}{\text{O}}

\allowdisplaybreaks[3]

\begin{document}

\begin{titlepage}
\renewcommand{\thefootnote}{\fnsymbol{footnote}}

\vspace*{1cm}

\centerline{\Large\textbf{Poisson--Lie $T$-plurality for WZW backgrounds}}%

\vspace{1.5cm}

\centerline{\large Yuho Sakatani}

\vspace{0.2cm}

\begin{center}
{\it Department of Physics, Kyoto Prefectural University of Medicine,}\\
{\it 1-5 Shimogamohangi-cho, Sakyo-ku, Kyoto, Japan}\\
{\small\texttt{yuho@koto.kpu-m.ac.jp}}
\end{center}

\vspace*{2mm}

\begin{abstract}
Poisson--Lie $T$-plurality constructs a chain of supergravity solutions from a Poisson--Lie symmetric solution. We study the Poisson--Lie $T$-plurality for supergravity solutions with $H$-flux, which are not Poisson--Lie symmetric but admit non-Abelian isometries, $\pounds_{v_a}g_{mn}=0$ and $\pounds_{v_a}H_3=0$ with $\pounds_{v_a}B_2\neq 0$. After introducing the general procedure, we study the Poisson--Lie $T$-plurality for two WZW backgrounds, the AdS$_3$ with $H$-flux and the Nappi--Witten background.
\end{abstract}

\thispagestyle{empty}
\end{titlepage}

\setcounter{footnote}{0}

\newpage

\tableofcontents

\newpage

\section{Introduction}

Abelian $T$-duality, which is a symmetry of string theory, is based on the existence of a set of commuting vector fields $v_a^m$ satisfying the Killing equations
\begin{align}
 \Lie_{v_a}g_{mn}=0\,,\qquad \Lie_{v_a}H_3=0\,.
\label{eq:isometry}
\end{align}
At least in supergravity, this symmetry continues to hold even when the Killing vector fields form a non-Abelian algebra $[v_a,\,v_b]=f_{ab}{}^c\,v_c$\,, and this is known as non-Abelian $T$-duality \cite{hep-th:9210021}. 
There is a further extension of the $T$-duality, called the Poisson--Lie (PL) $T$-duality \cite{hep-th:9502122,hep-th:9509095} or $T$-plurality \cite{hep-th:0205245}.
We can perform the PL $T$-duality when the background admits a set of vector fields $v_a^m$ satisfying the condition for the PL symmetry,
\begin{align}
 \Lie_{v_a}E_{mn} = - f_a{}^{bc}\,E_{mp}\,v^p_b\,v_c^q\,E_{qn} \qquad 
 \bigl(E_{mn}\equiv g_{mn}+B_{mn}\bigr)\,.
\label{eq:PL-symmetry}
\end{align}
Here $f_a{}^{bc}\,(=-f_a{}^{cb})$ are called the dual structure constants, and if they are absent, this condition reduces to the Killing equations, $\Lie_{v_a}g_{mn} =\Lie_{v_a}B_2=0$\,.
Then the PL $T$-duality reduces to the traditional non-Abelian $T$-duality. 
However, the condition $\Lie_{v_a}B_2=0$ is stronger than $\Lie_{v_a}H_3=0$ of Eq.~\eqref{eq:isometry}, and naively, not all of the traditional non-Abelian $T$-duality can be realized as a PL $T$-duality. 
In this short paper, we discuss how to perform the PL $T$-plurality in backgrounds with non-Abelian isometries, $\Lie_{v_a}g_{mn}=0$ and $\Lie_{v_a}H_3=0$ but $\Lie_{v_a}B_2\neq 0$\,. 

For the sake of clarity, let us comment on some of the related earlier works. 
There are at least two approaches to perform the PL $T$-duality in a $H$-fluxed background.\footnote{There is the third way that is based on the $\lambda$-deformation \cite{1312.4560}. For a target space of the $\lambda$-model, we can explicitly construct the generalized frame fields which produce the $H$-flux (see e.g.~\cite{1611.07978,1705.09304,1810.11446,1912.11036} for related techniques). Using these we can perform a PL-like $T$-duality. We thank Falk Hassler for valuable comments on this point.}
The first is taken in \cite{hep-th:9509123,1506.06233}, where Wess--Zumino--Witten (WZW) backgrounds are constructed as PL symmetric backgrounds satisfying the condition \eqref{eq:PL-symmetry} (see also \cite{1207.2304,1303.4069} for the supersymmetric extensions). 
For a given WZW background, how to find the set of vector fields $v_a^m$ and the dual structure constants $f_a{}^{bc}$ is quite non-trivial, but once we find these, we can perform the PL $T$-duality by following the standard procedure. 
The second approach is taken in \cite{1705.00458,1812.07664}, where the non-trivial $H$-flux is produced by the spectator fields and the internal part satisfies the usual Killing equations $\Lie_{v_a}E_{mn} = 0$\,. 
Namely, the WZW backgrounds are realized as PL symmetric backgrounds associated with semi-Abelian doubles (i.e., $f_a{}^{bc}=0$). 
Again, once we found such a realization, we can perform the PL $T$-duality by using the standard procedure of the PL $T$-duality with spectator fields.

Now we explain our approach. 
For convenience, we use the language of double field theory (DFT) \cite{hep-th:9302036,hep-th:9305073,hep-th:9308133,0904.4664}, whose flux formulation \cite{1304.1472} is especially useful here. 
In DFT, the supergravity fields in the NS--NS sector are packaged into the generalized metric and the DFT dilaton
\begin{align}
 \cH_{IJ} \equiv \begin{pmatrix} g_{mn} & (B\,g^{-1})_m{}^n \\ -(g^{-1}\,B)^m{}_n & g^{mn} \end{pmatrix},\qquad \Exp{-2d}\equiv \Exp{-2\Phi}\sqrt{\abs{\det g_{mn}}}\,,
\end{align}
where $I,J=1,\dotsc,2D$ and $m,n=1,\dotsc,D$\,. 
Similar to the standard setup to study the PL $T$-duality in DFT \cite{1707.08624}, we assume that the generalized metric has the form
\begin{align}
 \cH_{IJ}(x) = E_I{}^A(x)\,E_J{}^B(x)\,\hat{\cH}_{AB}\,,
\label{eq:H-ansatz}
\end{align}
where $\hat{\cH}_{AB}$ is constant and $E_I{}^A$ is the inverse of the generalized frame fields $E_A{}^I$ satisfying
\begin{align}
 [E_A,\,E_B]_D = - \cF_{AB}{}^C\,E_C\,. 
\label{eq:gen-Lie-algebra}
\end{align}
Here, $[\cdot,\cdot]_D$ denotes the D-bracket or the generalized Lie derivative in DFT and $\cF_{AB}{}^C$ are the structure constants with components
\begin{align}
 \cF_{abc}=H_{abc}\,,\qquad \cF_{ab}{}^c=\cF_b{}^c{}_a=\cF^c{}_{ab}=f_{ab}{}^c\,,\qquad
 \cF_a{}^{bc} = \cF^c{}_a{}^b=\cF^{bc}{}_a=0=\cF^{abc}\,,
\label{eq:cF-H}
\end{align}
under the decompositions, $\{{}_A\}=\{{}_a,\,{}^a\}$ and $\{{}^A\}=\{{}^a,\,{}_a\}$ ($A=1,\dotsc,2D$ and $a=1,\dotsc,D$). 
For the DFT dilaton, for simplicity, we assume the absence of the dilaton flux,
\begin{align}
 \cF_A \equiv E_A{}^I\,\cD_B E_I{}^B + 2\,\cD_A d = 0\qquad \bigl(\cD_A\equiv E_A{}^I\,\partial_I\bigr)\,.
\label{eq:dilaton-flux}
\end{align}
In the usual setup of the PL $T$-duality/plurality in DFT \cite{1707.08624,1810.11446,1903.12175}, the components of the structure constants are supposed to have the form
\begin{align}
 \cF_{ab}{}^c=-\cF_a{}^c{}_b=\cF^c{}_{ab}=f_{ab}{}^c\,,\quad
 \cF_a{}^{bc} = \cF^c{}_a{}^b=\cF^{bc}{}_a=f_a{}^{bc} \,,\quad
 \cF_{abc}=0=\cF^{abc}\,,
\label{eq:cF-DD}
\end{align}
which are the structure constants of the Drinfel'd double. 
However, as a solution generating technique, Eqs.~\eqref{eq:H-ansatz}, \eqref{eq:gen-Lie-algebra}, and \eqref{eq:dilaton-flux} are sufficient,\footnote{We thank Chris Blair for correspondence regarding this point.} and instead of \eqref{eq:cF-DD}, we rather suppose \eqref{eq:cF-H} in the initial background.
As usual, the equations of motion of DFT reduce to
\begin{align}
\begin{split}
 \hat{\cH}^{AD}\,\bigl(3\,\eta^{BE}\,\eta^{CF} - \hat{\cH}^{BE}\, \hat{\cH}^{CF}\bigr) \,\cF_{ABC}\,\cF_{DEF} &=0\,,
\\
 \bigl(\eta^{CE}\,\eta^{DF} - \hat{\cH}^{CE}\, \hat{\cH}^{DF}\bigr)\, \hat{\cH}^{G[A}\,\cF_{CD}{}^{B]}\,\cF_{EFT} &=0\,,
\end{split}
\label{eq:eom}
\end{align}
and they are covariant under the $\OO(D,D)$ transformation
\begin{align}
 \hat{\cH}_{AB}\to \hat{\cH}'_{AB} \equiv C_A{}^C\,C_B{}^D\,\hat{\cH}_{CD}\,,\qquad
 \cF_{AB}{}^C\to \cF'_{AB}{}^C \equiv C_A{}^D\,C_B{}^E\,(C^{-1})_F{}^C\,\cF_{DE}{}^F\,.
\label{eq:Onn}
\end{align}
Then, by finding a new set of generalized frame fields $E'_A{}^I$ satisfying $[E'_A,\,E'_B]_D=-\cF'_{AB}{}^C\,E'_C$ (but $E'_A\neq C_A{}^B\,E_B$), we can generate a new DFT solution $\cH'_{IJ} \equiv E'_I{}^A\,E'_J{}^B\,\hat{\cH}_{AB}$ and $d'$\,, which satisfies $\cF'_A\equiv E'_A{}^I\,\cD'_B E'_I{}^B + 2\,\cD'_A d'=0$\,. 

As concrete backgrounds satisfying Eqs.~\eqref{eq:H-ansatz}, \eqref{eq:gen-Lie-algebra}, and \eqref{eq:dilaton-flux}, we consider the target space of an (ungauged) WZW model. 
As we explain in section \ref{sec:WZW}, for any WZW model, we can explicitly construct the generalized frame fields $E_A{}^I$ and the DFT dilaton $d$ satisfying Eqs.~\eqref{eq:H-ansatz}--\eqref{eq:dilaton-flux}. 
There, we find a set of (generalized) Killing vector fields $(v_a^m,\,\tilde{v}_{am})$ satisfying
\begin{align}
 \Lie_{v_a}g_{mn}= 0\,,\qquad \Lie_{v_a} B_2 + \rmd \tilde{v}_a = 0 \,,\qquad
 [v_a,\,v_b]= f_{ab}{}^c\,v_c\,,
\label{eq:gen-iso}
\end{align}
which is similar to the setup of the traditional non-Abelian $T$-duality. 
Due to the presence of the 1-form fields $\tilde{v}_a$\,, this is different from the usual condition for the PL symmetry \eqref{eq:PL-symmetry} and the standard procedure of the PL $T$-duality cannot be applied. 
However, if the $2D$-dimensional Lie algebra $[T_A,\,T_B]=\cF_{AB}{}^C\,T_C$ with $\cF_{AB}{}^C$ given in \eqref{eq:cF-H} can be decomposed into two maximally isotropic subalgebras, we can construct the dual geometry by means of the PL $T$-plurality. 
Namely, if we find an $\OO(D,D)$ transformation \eqref{eq:Onn} such that the new structure constants $\cF'_{AB}{}^C$ have the form of Eq.~\eqref{eq:cF-DD}, the $2D$-dimensional algebra becomes a Drinfel'd double.
For a Drinfel'd double, we can systematically construct the dual fields $E'_A{}^I$ and $d'$ satisfying Eqs.~\eqref{eq:gen-Lie-algebra} and \eqref{eq:dilaton-flux}.
Then the resulting background should be a solution of DFT as long as the original WZW background is a solution of DFT. 
This is the idea of the solution generation method proposed in this paper. 

As concrete examples of the WZW model, we consider the $\SL(2)$ WZW model and the Nappi--Witten (NW) model \cite{hep-th:9310112}. 
For the $\SL(2)$ WZW model, we consider two approaches for the PL $T$-plurality: the one studied in \cite{hep-th:9509123} and our approach. 
We find that the two approaches are based on different six-dimensional Drinfel'd doubles. 
For each of the Drinfel'd doubles, there are several inequivalent decompositions into the pair of algebras $\{f_{ab}{}^c,\,f_c{}^{ab}\}$\,, called the Manin triples, and we construct a background for each of the Manin triples. 
In general, each Manin triple corresponds to a different target geometry but it turns out that many of the Manin triples correspond to the same AdS$_3$ solution with $H$-flux. 
This kind of self-duality under the PL $T$-duality has been observed in \cite{hep-th:9609112,1812.07664} and this may represent the high symmetry of the AdS$_3$ background. 
We also find that some Manin triples correspond to a flat space with a linear dilaton and some Manin triples correspond to a known AdS$_3$ solution of the generalized supergravity equations of motion \cite{1511.05795,1605.04884}. 

For the NW model, the Drinfel'd double is eight-dimensional and the Manin triples have not been classified. 
We find several inequivalent Manin triples and construct the corresponding backgrounds. 
Again, we find that two Manin triples correspond to the same NW background. 
Other Manin triples correspond to a kind of $T$-fold \cite{hep-th:0406102} which does not allow for the description in terms of the standard fields in the NS--NS sector. 
We also perform the Yang--Baxter (YB) deformation \cite{hep-th:0210095} of the NW background by using a classical $r$-matrix satisfying the modified classical YB equations (CYBE). 
As a result, we find a one-parameter family of solutions which contains the NW background and the flat Minkowski spacetime as specific cases. 

This paper is organized as follows. 
In section \ref{sec:WZW}, we explain how to construct the generalized frame fields $E_A{}^I$ and the DFT dilaton $d$ satisfying Eqs.~\eqref{eq:gen-Lie-algebra} and \eqref{eq:dilaton-flux} in WZW backgrounds. 
We also explain the procedure of the PL $T$-plurality. 
In section \ref{sec:SL2}, we study the PL $T$-plurality of the SL(2) WZW model. 
In section \ref{sec:SL2}, we study the PL $T$-plurality and the YB deformation of the NW background. 
Section \ref{sec:conclusion} is devoted to conclusions and discussions. 

\section{WZW background}
\label{sec:WZW}

Let us consider a group $G$ associated with a Lie algebra, $[T_a,\,T_b]=f_{ab}{}^c\,T_c$\,.
We define the left-/right-invariant 1-forms and their duals as
\begin{align}
 \ell \equiv g^{-1} \rmd g \equiv \ell^a_m \,\rmd x^m\,T_a\,, \quad
 r \equiv \rmd g\,g^{-1} \equiv r^a_m \,\rmd x^m\,T_a\,,\quad
 v_a^m\,\ell_m^b = \delta_a^b\,,\quad
 e_a^m\,r_m^b = \delta_a^b\,.
\end{align}
Then the WZW model can be defined as
\begin{align}
 S = \frac{1}{4\pi\alpha'} \int_{\Sigma} g_{mn} \, \rmd x^m \wedge *\rmd x^n + \frac{1}{2\pi \alpha'} \int_B H_3\qquad (\partial B=\Sigma)\,,
\end{align}
where the metric and the 3-form field strength are
\begin{align}
\begin{split}
 g_{mn} &\equiv \hat{g}_{ab}\,\ell^a_m\,\ell^b_n = \hat{g}_{ab}\,r^a_m\,r^b_n\,,
\\
  H_3 &\equiv \frac{1}{3!}\,f_{abc}\,\ell^a\wedge \ell^b\wedge\ell^c 
  = \frac{1}{3!}\,f_{abc}\,r^a\wedge r^b\wedge r^c \,.
\end{split}
\label{eq:WZW-bg}
\end{align}
Here $\hat{g}_{ab}$ is a non-degenerate invariant metric and $f_{abc}\equiv f_{ab}{}^d\,\hat{g}_{dc}$ is totally antisymmetric. 

If we introduce a 2-form field $B_2$ satisfying $\rmd B_2=H_3$\,, we can check the generalized Killing vector fields $V_a{}^I\equiv (v_a^m,\,\tilde{v}_{am}\equiv E_{mn}\,v_a^n\,\rmd x^m)$ satisfy Eq.~\eqref{eq:gen-iso} for any choice of $B_2$\,. 
In general, $\tilde{v}_{am}$ does not vanish and this background is not PL symmetric, i.e., we cannot perform the PL $T$-duality. 
Now we define the generalized frame fields and a constant matrix as
\begin{align}
 E_A{}^I \equiv \begin{pmatrix} e_a^m & -e_a^n\,B_{nm} \\ 0 & r^a_m \end{pmatrix}, \qquad
 \hat{\cH}_{AB}\equiv \begin{pmatrix} \hat{g}_{ab} & 0 \\ 0 & \hat{g}^{ab}
\end{pmatrix},
\label{eq:E-cH}
\end{align}
and then the generalized metric for the WZW background \eqref{eq:WZW-bg} is expressed as
\begin{align}
 \cH_{IJ}(x) = E_I{}^A(x)\,E_J{}^B(x)\,\hat{\cH}_{AB}\,.
\end{align}
Using $e_a^m\,e_b^n\,e_c^p\,H_{mnp}=f_{abc}$\,, we can easily check that the frame fields $\{E_A\}=\{E_a,\,E^a\}$ satisfy the algebra \eqref{eq:gen-Lie-algebra} with the fluxes $\cF_{AB}{}^C$ given by Eq.~\eqref{eq:cF-H} and $H_{abc}=f_{abc}$,\footnote{If a $B$-field satisfies $\rmd B_2=\frac{1}{3!}\,H_{abc}\,r^a\wedge r^b\wedge r^c$ for a general skew-symmetric constants $H_{abc}$, the $E_A{}^I$ defined in Eq.~\eqref{eq:E-cH} satisfy the algebra \eqref{eq:alg-H=f} with $f_{abc}$ replaced by $H_{abc}$\,. In such a general case, $\cH_{IJ}=E_I{}^A\,E_J{}^B\,\hat{\cH}_{AB}$ does not describe the target space of a WZW model, but we can still perform the PL $T$-plurality. For the PL $T$-plurality, $\hat{\cH}_{AB}$ also can be an arbitrary constant $\OO(D,D)$ matrix and is not restricted to have the form \eqref{eq:E-cH}.}
\begin{align}
\begin{split}
 [E_a,\, E_b]_D &= - f_{ab}{}^c\,E_c - f_{abc}\,E^{c}\,,\quad
 [E^a,\,E^b]_D = 0\,,
\\
 [E_a,\, E^b]_D &= - [E^b,\,E_a]_D = f_{ac}{}^b\,E^{c} \,.
\end{split}
\label{eq:alg-H=f}
\end{align}
Here the D-bracket or the generalized Lie derivative in DFT is defined as
\begin{align}
 [V,\,W]_D^I=V^J\,\partial_J W^I -\bigl(\partial_J V^I - \partial^I V_J\bigr)\,W^J\,,
\end{align}
where $(\partial_I)=(\frac{\partial}{\partial x^m},\,\frac{\partial}{\partial\tilde{x}_m})$ and the indices $\{{}^I\}=\{{}^m,\,{}_m\}$ and $\{{}_I\}=\{{}_m,\,{}^m\}$ are raised or lowered with the $\OO(D,D)$-invariant metric
\begin{align}
 \eta_{IJ} \equiv \begin{pmatrix} 0 & \delta_m^n \\ \delta^m_n & 0\end{pmatrix},\qquad
 \eta^{IJ} \equiv \begin{pmatrix} 0 & \delta^m_n \\ \delta_m^n & 0\end{pmatrix}.
\end{align}
Regarding the DFT dilaton, the requirement \eqref{eq:dilaton-flux} reduces to
\begin{align}
 -2\,\partial_m d = e_b^n\,\partial_n r_m^b = -\partial_n e_b^n\,r_m^b \,,
\end{align}
under our assumption $\frac{\partial}{\partial\tilde{x}_m}=0$\,. 
By using $\partial_n e_b^n = \ell^a_m\, e_b^n\,\partial_n v_a^m = -e_b^n\,\partial_n \ln\abs{\det(\ell^a_m)}$\,, which follows from $\ell^a_m\,\Lie_{v_a} e_b^m=0$\,, the DFT dilaton can be found to have the form
\begin{align}
 \Exp{-2\,d} = \Exp{-2\,d_0} \abs{\det(\ell^a_m)}\,,
\end{align}
where $d_0$ is a constant. 
In our examples, we simply choose $d_0=0$ because the constant is not important in the equations of motion. 

Now let us perform the $\OO(D,D)$ transformation \eqref{eq:Onn} such that the new algebra $\cF'_{AB}{}^C$ is the Lie algebra of a Drinfel'd double. 
Then we parameterize the group element, for example, as $g(x)=\Exp{x^a\,T_a}$ and define the right-/left-invariant 1-forms and their duals. 
We also define
\begin{align}
 g^{-1}\,T_A\,g \equiv \begin{pmatrix} a_a{}^b & 0 \\ (\pi\,a)^{ab} & (a^{-1})_b{}^a \end{pmatrix} \qquad \bigl(\pi^{ab}=-\pi^{ba}\bigr)\,,
\end{align}
and construct new generalized frame fields and a DFT dilaton as
\begin{align}
 E_A{}^I(x) \equiv \begin{pmatrix} e_a^m & 0 \\ \pi^{ab}\,e_b^m & r^a_m \end{pmatrix},\qquad 
 \Exp{-2\,d(x)} = \Exp{-2\,d_0} \abs{\det(\ell^a_m)}\,.
\label{eq:E-d-PL}
\end{align}
They are known to satisfy the desired properties \eqref{eq:gen-Lie-algebra} and \eqref{eq:dilaton-flux} \cite{1707.08624,1810.11446}, and using these fields, we obtain the dual background, in the same way as the standard PL $T$-plurality.\footnote{When the dual algebra is non-unimodular ($f_a{}^{ab}\neq 0$), a PL symmetric background is known to satisfy the generalized supergravity equations of motion \cite{1511.05795,1605.04884} with the vector field $I=\frac{1}{2}\,f_b{}^{ba}\,v_a$ \cite{1810.11446}.}

In order to make a solution of ten-dimensional supergravity, we may consider a product of the WZW background and a certain space with coordinates $y^\mu$\,, called the spectator fields. 
In the examples to be studied in this paper, the generalized metric is given by a direct sum
\begin{align}
 \bm{\cH}_{\hat{I}\hat{J}}(x^m,\,y^\mu) = \begin{pmatrix} \cH_{IJ}(x) & 0 \\ 0 & \cH^s_{MN}(y) \end{pmatrix},
\end{align}
where $\hat{I}=1,\dotsc,20$ and $\{{}_M\}=\{{}_\mu,\,{}^\mu\}$ with $\mu=1,\dotsc,10-n$\,. The DFT dilaton is given by a product
\begin{align}
 \Exp{-2\,\bm{d}(x^m,\,y^\mu)} = \Exp{-2\,d(x)} \Exp{-2\,d_s(y)} .
\end{align}
The fields $\cH^s_{MN}(y)$ and $d_s(y)$ are invariant under the PL $T$-pluralities and only $\cH_{IJ}(x)$ and $d(x)$ are transformed in the following discussion. 

\section{SL(2) WZW model}
\label{sec:SL2}

Here we consider the SL(2) WZW model, where the target geometry can be identified with the AdS$_3$ with $H$-flux
\begin{align}
 \rmd s^2 = l^2\,\frac{\rmd z^2 - \rmd t^2 + \rmd x^2}{z^2}\,, \qquad
 H_3 = \pm \frac{l^2\,\rmd z\wedge \rmd t\wedge\rmd x}{z^3} = \pm 2\,l^{-1} * 1 \,.
\label{eq:ads3-H}
\end{align}
The Riemann curvature tensors is $R_{mnpq}=-l^{-2}\,(g_{mp}\,g_{nq}-g_{mq}\,g_{np})$ and the Ricci scalar is $R=-6/l^2$\,. 
This becomes a ten-dimensional supergravity solution by adding spectator fields (which are not affected by $T$-dualities)
\begin{align}
\begin{split}
 \rmd s^2_{\text{S}^3\times T^4} &= \tfrac{l^2}{4}\,\bigl[\rmd \theta^2 + \sin^2\theta\,\rmd \phi^2 + \bigl(\rmd \psi + \cos\theta\,\rmd \phi\bigr)^2\bigr] + \rmd y_1^2+ \rmd y_2^2+ \rmd y_3^2+ \rmd y_4^2\,,
\\
 H^{(\text{S}^3\times T^4)}_3 &= \tfrac{l^2}{4}\sin\theta\,\rmd\theta\wedge\rmd\phi\wedge\rmd\psi\,.\qquad \Exp{-2\,d_s(y)}=(l/2)^3\sin\theta\,.
\end{split}
\label{eq:external}
\end{align}
Since $\Exp{-2\,d_s(y)}$ is just the volume element, the dilaton $\Phi$ can be found as $\Exp{-2\,\Phi}=\frac{\Exp{-2\,d(x)}}{\sqrt{\abs{\det g_{mn}}}}$\,. 

In \cite{hep-th:9509123}, this WZW background was reproduced as a PL symmetric background by using a six-dimensional Drinfel'd double. 
Six-dimensional Drinfel'd doubles and the Manin triples have been classified in \cite{math/0202210} and it is known that there are 22 Drinfel'd doubles. 
According to this classification, the Drinfel'd double used in \cite{hep-th:9509123} is called DD2 (see \cite{hep-th:0403164} for the notation). 
This Drinfel'd double can be decomposed into four Manin triples \cite{math/0202210}
\begin{align}
 \text{DD2:}\quad(\bm{5.i}|\bm{8}|b) \cong (\bm{8}|\bm{5.i}|b)\cong
 (\bm{6_0}|\bm{5.iii}|b) \cong (\bm{5.iii}|\bm{6_0}|b) \qquad (b>0)\,,
\label{eq:DD2-orbit}
\end{align}
and it was found that the SL(2) WZW background is associated with $(\bm{5.i}|\bm{8}|b)$ \cite{hep-th:9509123}. 
Then the PL $T$-dual model, whose Manin triple is $(\bm{8}|\bm{5.i}|b)$, was found to be a constrained sigma model \cite{hep-th:9509123}. 
In this section, we find that the target geometry of this dual model is a kind of non-Riemannian backgrounds (see \cite{1307.8377,1707.03713,1909.10711} for details of non-Riemannian backgrounds).
Then we further perform PL $T$-plurarities and obtain the backgrounds associated with the other two Manin triples, $(\bm{6_0}|\bm{5.iii}|b)$ and $(\bm{5.iii}|\bm{6_0}|b)$. 

After completing the orbit of DD2, we consider the PL $T$-pluralities based on our approach. 
We start with a flux algebra with the SL(2) algebra $f_{ab}{}^c$ and the $H$-flux $H_{abc}\neq 0$\,. 
Then we identify that this six-dimensional Lie algebra corresponds to a Drinfel'd double called DD7. 
The DD7 can be decomposed into six Manin-triples \cite{hep-th:9509123}
\begin{align}
 \text{DD7:}\quad (\bm{7_0}|\bm{4}|b) \cong (\bm{4}|\bm{7_0}|b)\cong
 (\bm{4}|\bm{2.iii}|b) \cong (\bm{2.iii}|\bm{4}|b)\cong
 (\bm{6_0}|\bm{4.i}|-b) \cong (\bm{4.i}|\bm{6_0}|-b) \,,
\label{eq:DD7-orbit}
\end{align}
and we identify the corresponding backgrounds. 

\subsection{PL $T$-plurality for DD2}

Here we review the result of \cite{hep-th:9509123} and then complete the orbit of DD2 described in Eq.~\eqref{eq:DD2-orbit}.

\subsubsection{Manin triple \texorpdfstring{$(\bm{5.i}|\bm{8}|1)$}{(5.i|8|1)}}

The Lie algebra of the Drinfel'd double considered in \cite{hep-th:9509123} has the following structure constants
\begin{align}
 f_{12}{}^2 = 1\,,\quad 
 f_{13}{}^3 = 1\,,\quad 
 f_2{}^{12} = 1\,,\quad 
 f_3{}^{13} = -1\,,\quad 
 f_1{}^{23} = 2\,.
\label{eq:9509123-alg}
\end{align}
If we consider a redefinition,
\begin{align}
 T_a \to \Lambda_a{}^b\,T_b\,,\qquad
 T^a \to (\Lambda^{-1})_b{}^a\,T^b\,,\qquad
 \Lambda_a{}^b = {\footnotesize\begin{pmatrix}
 -1 & 0 & 0 \\
 0 & 2 & \frac{1}{2} \\
 0 & 2 & -\frac{1}{2} \end{pmatrix}}, 
\end{align}
the structure constants become
\begin{align}
 f_{12}{}^2 = -1\,,\quad
 f_{13}{}^3 = -1\,,\quad
 f_3{}^{12} = -1\,,\quad
 f_1{}^{23} = 1\,,\quad
 f_2{}^{13} = -1\,.
\end{align}
Therefore, this Manin triple is isomorphic to $(\bm{5.i}|\bm{8}|b=1)$ of \cite{math/0202210}. 

Using the algebra \eqref{eq:9509123-alg} and a parameterization $g=\Exp{x\,T_1}\Exp{y\,T_2+z\,T_3}$, we obtain
\begin{align}
 r_m{}^a &= {\footnotesize\begin{pmatrix}
 1 & 0 & 0 \\
 0 & \Exp{x} & 0 \\
 0 & 0 & \Exp{x} \end{pmatrix}},\qquad
 \ell_m{}^a = {\footnotesize\begin{pmatrix}
 1 & y & z \\
 0 & 1 & 0 \\
 0 & 0 & 1 \end{pmatrix}},
\\
 \pi^{ab} &= 
 {\footnotesize\begin{pmatrix}
 0 & -\Exp{x} y & \Exp{x}z \\
 \Exp{x} y & 0 & 1-\Exp{2x} (y\,z+1) \\
 -\Exp{x} z & \Exp{2x}(y\,z+1)-1 & 0 \end{pmatrix}}.
\end{align}
By following \cite{hep-th:9509123}, we introduce the constant matrix as
\begin{align}
 \hat{\cH}_{AB} = {\footnotesize\begin{pmatrix}
 1 & 0 & 0 & 0 & 0 & 0 \\
 0 & 0 & 0 & 0 & -1 & 0 \\
 0 & 0 & 0 & 0 & 0 & 1 \\
 0 & 0 & 0 & 1 & 0 & 0 \\
 0 & -1 & 0 & 0 & 0 & 2 \\
 0 & 0 & 1 & 0 & 2 & 0\end{pmatrix}},
\end{align}
and then compute the generalized metric $\cH_{IJ}(x)=E_I{}^A\,E_J{}^B\,\hat{\cH}_{AB}$\,, where $E_A{}^I$ is defined in Eq.~\eqref{eq:E-d-PL}.
Then the three-dimensional part of the supergravity fields are found as
\begin{align}
 g_{mn} = \begin{pmatrix}
 1+y\,z & \frac{z}{2} & \frac{y}{2} \\
 \frac{z}{2} & 0 & \frac{1}{2} \\
 \frac{y}{2} & \frac{1}{2} & 0\end{pmatrix} ,\qquad
 B_{mn} = \begin{pmatrix}
 0 & \frac{z}{2} & -\frac{y}{2} \\
 -\frac{z}{2} & 0 & -\frac{1}{2} \\
 \frac{y}{2} & \frac{1}{2} & 0\end{pmatrix}.
\end{align}
The DFT dilaton is trivial $\Exp{-2\,d(x)}=\Exp{-2\,d_0}\abs{\det(\ell_m^a)}=\Exp{-2\,d_0}=1$\,, and the dilaton $\Phi$ is just a constant. 
This background is a conformally flat Einstein space with $R=-6$ and the $H$-flux is $H_3=\rmd x\wedge \rmd y\wedge \rmd z = *2$\,. 
Namely, at least locally, this is precisely the $H$-fluxed AdS$_3$ background \eqref{eq:ads3-H} with unit AdS radius $l=1$\,. 

We can easily check that the condition for the PL symmetry \eqref{eq:PL-symmetry} is satisfied by the left-invariant vector fields $v_a^m$ and the structure constants \eqref{eq:9509123-alg}. 
We can also check that these vector fields $v_a^m$ do not generate the isometries of the target space, $\Lie_{v_a}g_{mn}\neq 0$ and $\Lie_{v_a}H_3\neq 0$\,, unlike the usual setup of the traditional (non-)Abelian $T$-duality. 

\subsubsection{Manin triple \texorpdfstring{$(\bm{8}|\bm{5.i}|1)$}{(8|5.i|1)}}

Considering the PL $T$-duality $T_a\leftrightarrow T^a$, namely, the $\OO(3,3)$ transformation \eqref{eq:Onn} with
\begin{align}
 C_A{}^B = \begin{pmatrix} \bm{0}_3 & \bm{1}_3 \\ \bm{1}_3 & \bm{0}_3 \end{pmatrix},
\end{align}
the algebra is mapped to a Manin triple $(\bm{8}|\bm{5.i}|1)$\,, where $f_{abc}{}^c$ and $f_c{}^{ab}$ are swapped. 
The constant metric becomes
\begin{align}
 \hat{\cH}_{AB} = {\footnotesize\begin{pmatrix}
 1 & 0 & 0 & 0 & 0 & 0 \\
 0 & 0 & 2 & 0 & -1 & 0 \\
 0 & 2 & 0 & 0 & 0 & 1 \\
 0 & 0 & 0 & 1 & 0 & 0 \\
 0 & -1 & 0 & 0 & 0 & 0 \\
 0 & 0 & 1 & 0 & 0 & 0 \end{pmatrix}}.
\end{align}
By using a parameterization $g=\Exp{z\,T_3}\Exp{y\,T_2}\Exp{x\,T_1}$, we find
\begin{align}
 r_m{}^a &= {\footnotesize\begin{pmatrix}
 1+2\,y\,z & -y & y\,z^2+z \\
 -2\,z & 1 & -z^2 \\
 0 & 0 & 1\end{pmatrix}} ,\qquad
 \ell_m{}^a = {\footnotesize\begin{pmatrix}
 1 & 0 & 0 \\
 0 & \Exp{-x} & 0 \\
 -2\, y & -\Exp{-x} y^2 & \Exp{x} \end{pmatrix}},
\\
 \pi^{ab}&={\footnotesize\begin{pmatrix}
 0 & -y & -z\,(1+y\,z) \\
 y & 0 & y\,z \\
 z\,(1+y\,z) & -y\,z & 0 \end{pmatrix}},
\end{align}
and the generalized metric and the DFT dilaton become
\begin{align}
 \cH_{IJ}(x)= {\footnotesize\begin{pmatrix}
 1 & 0 & -2\,y & 0 & 0 & 0 \\
 0 & 0 & 2 & 0 & -1 & 0 \\
 -2\,y & 2 & 0 & 2\,y & 2\,y^2 & 1 \\
 0 & 0 & 2\,y & 1 & 0 & 0 \\
 0 & -1 & 2\,y^2 & 0 & 0 & 0 \\
 0 & 0 & 1 & 0 & 0 & 0\end{pmatrix}} ,\qquad d(x)= 0\,.
\end{align}
Since the dual algebra is non-unimodular, this background is a solution of the generalized supergravity equations of motion with the vector field $I=\frac{1}{2}\,f_b{}^{ba}\,v_a=-\partial_x$\,. 
In DFT, the constant vector field $I$ can be included into the DFT dilaton as $d(x)=I^m\,\tilde{x}_m = -\tilde{x}$ \cite{1703.09213}, and then we find that this background is a solution of DFT (by adding the spectator fields). 

Since the matrix $\cH^{mn}$ is degenerate, the standard fields $\{g_{mn},\,B_{mn},\,\Phi\}$ are not defined and this is an example of non-Riemannian backgrounds \cite{1307.8377}.
If we parameterize $\cH_{IJ}(x)$ as
\begin{align}
 \cH_{IJ}(x)= \begin{pmatrix} G_{mn} & G_{mq}\,\beta^{qn} \\ -\beta^{mq}\,G_{qn} & G^{mn} - \beta^{mp}\,G_{pq}\,\beta^{qn} \end{pmatrix},
\end{align}
we find
\begin{align}
 G_{mn} = {\footnotesize\begin{pmatrix} 1 & 0 & -2\,y \\
 0 & 0 & 2 \\
 -2\,y & 2 & 0 \end{pmatrix}},\qquad
 \beta^{mn} = {\footnotesize\begin{pmatrix} 0 & -y & 0 \\
 y & 0 & \frac{1}{2} \\
 0 & -\frac{1}{2} & 0 \end{pmatrix}},\qquad d(x)=-\tilde{x}\,.
\end{align}
It turns out that the (open-string) metric $G_{mn}$ is the AdS$_3$ with the radius $l=2$\,, and the $\beta$-field produces a constant $Q$-flux $Q_y{}^{xy}\equiv \partial_y\beta^{xy}=-1$ with a non-zero trace. 
This will be the target geometry of the constrained sigma model studied in \cite{hep-th:9509123}.\footnote{In \cite{1307.8377}, the string sigma model in a general non-Riemannian background is shown to be constrained by a certain self-duality relation (or a chirality constraint), and this seems to be consistent with the result of \cite{hep-th:9509123}.} 

We can remove the $\tilde{x}$-dependence of the dilaton, by performing an Abelian $T$-duality along the $x$-direction. 
We then find
\begin{align}
 G_{mn}(x) = {\footnotesize\begin{pmatrix} 1 & 0 & 2\,y \\
 0 & 0 & 2 \\
 2\,y & 2 & 0 \end{pmatrix}},\qquad
 \beta^{mn}(x) = {\footnotesize\begin{pmatrix} 0 & y & 0 \\
 -y & 0 & \frac{1}{2} \\
 0 & -\frac{1}{2} & 0 \end{pmatrix}},\qquad d(x)=-x\,,
\end{align}
which is again the AdS$_3$ geometry with a constant $Q$-flux. 
Moreover, in order to describe this background in the usual supergravity fields, let us consider a further constant $\OO(3,3)$ transformation (or a constant $B$-shift)
\begin{align}
 \cH_{IJ}\to O_I{}^K\,O_J{}^L\,\cH_{KL}\,,\qquad
 O_I{}^J= \begin{pmatrix} \delta_m^n & 0 \\ b_{mn} & \delta^m_n \end{pmatrix},\qquad 
 b_{mn} = {\footnotesize\begin{pmatrix}
 0 & 0 & \eta \\
 0 & 0 & 0 \\
 -\eta & 0 & 0 
\end{pmatrix}}.
\end{align}
In terms of $\{G_{mn},\,\beta^{mn}\}$, we find
\begin{align}
 G_{mn}(x) = {\footnotesize\begin{pmatrix} 1 & 0 & 2\,y \\
 0 & 0 & 2 \\
 2\,y & 2 & 0 \end{pmatrix}},\qquad
 \beta^{mn}(x) = {\footnotesize\begin{pmatrix} 0 & y & -\eta \\
 -y & 0 & \frac{1}{2} \\
 \eta & -\frac{1}{2} & 0 \end{pmatrix}},\qquad d(x)=-x\,,
\end{align}
and only the $\beta$-field gets a constant shift. 
In terms of the usual (closed-string) metric $g_{mn}$ and the $B$-field, the non-zero $\eta$ is crucial and we find a solution of the usual supergravity
\begin{align}
\begin{split}
 g_{mn} &= {\footnotesize\begin{pmatrix} 0 & \frac{1}{2\,\eta\,y^2} & 0 \\
 \frac{1}{2\,\eta\,y^2} & \frac{1}{y^2} & \frac{2\,\eta\,y-1}{2\,\eta^2\,y^2} \\
 0 & \frac{2\,\eta\,y-1}{2\,\eta ^2\,y^2} & \frac{1}{\eta^2} \end{pmatrix}},\qquad
 B_{mn} = {\footnotesize\begin{pmatrix} 0 & \frac{1}{2\,\eta\,y^2} & -\frac{1}{\eta} \\
 -\frac{1}{2\,\eta\,y^2} & 0 & \frac{1-2\,\eta\,y}{2\,\eta^2\,y^2} \\
 \frac{1}{\eta} & \frac{2\,\eta\,y-1}{2\,\eta^2\,y^2} & 0 \end{pmatrix}},
\\
\Exp{-2\Phi} &=2\,\eta^2\,y^2\Exp{2\,x} \,.
\end{split}
\label{eq:flat-dilaton}
\end{align}
Interestingly, this is a flat metric without $H$-flux for an arbitrary value of $\eta\,(\neq0)$, and the dilaton satisfies $\nabla_m\partial_n\Phi=0$ and $g^{mn}\,\partial_m\Phi\,\partial_n\Phi=1$\,. 
Thus, we can find a certain coordinate transformation which makes this solution a flat space with a (spacelike) linear dilaton $\Phi=\pm x$\,. 

\subsubsection{Manin triple \texorpdfstring{$(\bm{5.iii}|\bm{6_0}|1)$}{(5.iii|6\textzeroinferior|1)}}

From the algebra \eqref{eq:9509123-alg}, we consider a redefinition of generators $T_A\to C_A{}^B\,T_B$ with
\begin{align}
 C_A{}^B = {\footnotesize\begin{pmatrix}
 0 & -2 & \frac{1}{2} & 1 & -\frac{1}{4} & -1 \\
 1 & 0 & 0 & 0 & \frac{1}{4} & -1 \\
 0 & -2 & -\frac{1}{2} & 0 & \frac{1}{4} & -1 \\
 \frac{1}{2} & 0 & 0 & 0 & -\frac{1}{8} & \frac{1}{2} \\
 0 & 1 & -\frac{1}{4} & \frac{1}{2} & \frac{1}{8} & \frac{1}{2} \\
 0 & 0 & 0 & 0 & -\frac{1}{4} & -1 \end{pmatrix}}.
\end{align}
We then obtain a new Manin triple with
\begin{align}
 f_{23}{}^2 = -1\,,\quad
 f_{13}{}^1 = -1\,,\quad
 f_1{}^{23} = 1\,,\quad
 f_2{}^{13} = 1\,,
\end{align}
which is called $(\bm{5.iii}|\bm{6_0}|1)$\,. 
Under the transformation, the constant matrix is transformed as
\begin{align}
 \hat{\cH}_{AB}={\footnotesize\begin{pmatrix}
 0 & 0 & 0 & 0 & 1 & 0 \\
 0 & 0 & 0 & 1 & 0 & 0 \\
 0 & 0 & 1 & 0 & 0 & 0 \\
 0 & 1 & 0 & 0 & 0 & 0 \\
 1 & 0 & 0 & 0 & 0 & 0 \\
 0 & 0 & 0 & 0 & 0 & 1\end{pmatrix}}.
\end{align}
Considering a parameterization $g=\Exp{x\,T_1+y\,T_2}\Exp{z\,T_3}$, we find
\begin{align}
 r_m{}^a &= {\footnotesize\begin{pmatrix}
 1 & 0 & 0 \\
 0 & 1 & 0 \\
 -x & -y & 1 \end{pmatrix}} ,\qquad
 \ell_m{}^a = {\footnotesize\begin{pmatrix}
 \Exp{-z} & 0 & 0 \\
 0 & \Exp{-z} & 0 \\
 0 & 0 & 1 \end{pmatrix}},
\\
 \pi^{ab}&={\footnotesize\begin{pmatrix}
 0 & -\frac{x^2-y^2}{2} & -y \\
 \frac{x^2-y^2}{2} & 0 & -x \\
 y & x & 0 \end{pmatrix}} .
\end{align}
The generalized metric and the DFT dilaton can be found as
\begin{align}
 \cH_{IJ}(x)={\footnotesize\begin{pmatrix}
 0 & 0 & 0 & 0 & 1 & 0 \\
 0 & 0 & 0 & 1 & 0 & 0 \\
 0 & 0 & 1 & 0 & 0 & 0 \\
 0 & 1 & 0 & 0 & 0 & 0 \\
 1 & 0 & 0 & 0 & 0 & 0 \\
 0 & 0 & 0 & 0 & 0 & 1\end{pmatrix}},\qquad \Exp{-2\,d(x)} = \abs{\det\ell^a_m} = \Exp{-2\,z} .
\end{align}
This is again a non-Riemannian background, but if we perform a factorized $T$-duality along the $x$-direction (which is a symmetry of string theory), we get a Riemannian background
\begin{align}
 \rmd s^2 = 2\,\rmd x\,\rmd y+\rmd z^2 \,,\quad B_2 =0\,,\quad \Phi = z\,.
\label{eq:flat-linear-dilaton}
\end{align}
This (together with the spectator fields) is a solution of the ten-dimensional supergravity. 
Then, we again found that the AdS$_3$ background with $H$-flux is related to the flat space with a linear dilaton through a PL $T$-plurality (and the usual $T$-duality). 

\subsubsection{Manin triple \texorpdfstring{$(\bm{6_0}|\bm{5.iii}|1)$}{(6\textzeroinferior|5.iii|1)}}

We can realize the Manin triple $(\bm{6_0}|\bm{5.iii}|1)$ by considering the PL $T$-duality from the previous example. 
Again we consider a parameterization $g=\Exp{x\,T_1+y\,T_2}\Exp{z\,T_3}$ and then obtain
\begin{align}
 r_m{}^a = {\footnotesize\begin{pmatrix}
 1 & 0 & 0 \\
 0 & 1 & 0 \\
 y & x & 1 \end{pmatrix}} ,\qquad
 \ell_m{}^a = {\footnotesize\begin{pmatrix}
 \cosh z & \sinh z & 0 \\
 \sinh z & \cosh z & 0 \\
 0 & 0 & 1 \end{pmatrix}},
\qquad
 \pi^{ab} ={\footnotesize\begin{pmatrix}
 0 & \frac{x^2-y^2}{2} & x \\
 -\frac{x^2-y^2}{2} & 0 & y \\
 -x & -y & 0 \end{pmatrix}} .
\end{align}
The generalized metric and the DFT dilaton become
\begin{align}
 \cH_{IJ}(x)={\footnotesize\begin{pmatrix}
 0 & 0 & 0 & 0 & 1 & 0 \\
 0 & 0 & 0 & 1 & 0 & 0 \\
 0 & 0 & 1 & 0 & 0 & 0 \\
 0 & 1 & 0 & 0 & 0 & 0 \\
 1 & 0 & 0 & 0 & 0 & 0 \\
 0 & 0 & 0 & 0 & 0 & 1\end{pmatrix}},\qquad d(x) = 0\,.
\end{align}
Since the dual algebra is non-unimodular, the vector field $I = -\partial_z$ is needed to satisfy the generalized supergravity equations of motion. 
Again, under a factorized $T$-duality along the $x$-direction, this non-Riemannian background is mapped to a flat Riemannian background
\begin{align}
 \rmd s^2 = 2\,\rmd x\,\rmd y+\rmd z^2 \,,\quad B_2 =0\,,\quad \Phi =0\,,\quad I = -\partial_z\,.
\end{align}
Moreover, we can include $I$ into the DFT dilaton as $d(x)=-\tilde{z}$\,, and the by performing the formal $T$-duality along the $z$-direction (which interchanges $z$ and $\tilde{z}$), we get
\begin{align}
 \rmd s^2 = 2\,\rmd x\,\rmd y+\rmd z^2 \,,\quad B_2 =0\,,\quad \Phi = - z\,.
\end{align}
Then, again, we obtained the flat space with a linear dilaton, although the sign of the dilaton is changed. 

\subsection{PL $T$-plurality for DD7}

Now we consider the PL $T$-plurality based on our approach.
We consider the $\SL(2)$ algebra
\begin{align}
 f_{12}{}^3=-1\,,\qquad
 f_{13}{}^1=-1\,,\qquad
 f_{23}{}^2=1\,,
\end{align}
and denote an invariant metric as
\begin{align}
 \hat{g}_{ab} = {\footnotesize\begin{pmatrix} 0 & -1 & 0 \\ -1 & 0 & 0 \\ 0 & 0 & 1 \end{pmatrix}}.
\label{eq:sl2-invg}
\end{align}
Then we find the non-vanishing components of $f_{abc}\equiv f_{ab}{}^d\,\hat{g}_{dc}$ as $f_{123}=-1$\,. 

Using the coordinates $x^m=(z,t,x)$ and a parameterization
\begin{align}
 g = \Exp{\sqrt{2}\,(t - x)\,T_2} \Exp{-2\ln z\,T_3} \Exp{\sqrt{2}\,(t + x)\,T_1},
\end{align}
we obtain 
\begin{align}
 r_m{}^a &= {\footnotesize\begin{pmatrix}
 0 & -\frac{2 \sqrt{2} (t-x)}{z} & -\frac{2}{z} \\
 \frac{\sqrt{2}}{z^2} & \frac{\sqrt{2}\,[(t-x)^2+z^2]}{z^2} & \frac{2 (t-x)}{z^2} \\
 \frac{\sqrt{2}}{z^2} & \frac{\sqrt{2}\,[(t-x)^2-z^2]}{z^2} & \frac{2 (t-x)}{z^2} \end{pmatrix}} ,\qquad
 \ell_m{}^a = {\footnotesize\begin{pmatrix}
 -\frac{2 \sqrt{2} (t+x)}{z} & 0 & -\frac{2}{z} \\
 \frac{\sqrt{2}\,[(t+x)^2+z^2]}{z^2} & \frac{\sqrt{2}}{z^2} & \frac{2 (t+x)}{z^2} \\
 -\frac{\sqrt{2}\,[(t+x)^2-z^2]}{z^2} & -\frac{\sqrt{2}}{z^2} & -\frac{2 (t+x)}{z^2} \end{pmatrix}}.
\end{align}
Then the metric and the $H$-flux of the WZW background are found as
\begin{align}
 \rmd s^2 = 4\,\frac{\rmd z^2 - \rmd t^2 + \rmd x^2}{z^2}\,, \qquad
 H_3 = \frac{1}{3!}\,f_{abc}\,r^a\wedge r^b\wedge r^c 
 = -\frac{8\,\rmd z\wedge \rmd t\wedge\rmd x}{z^3}\,.
\end{align}
Here $\Lie_{v_a}g_{mn}=0$ and $\Lie_{v_a}H_3=0$ are satisfied, but we find $\Lie_{v_2}B_2\neq 0$ for $B_2 = \frac{4}{z^2}\, \rmd t \wedge\rmd x$\,.
The AdS radius is $l=2$ and we add spectator fields \eqref{eq:external} with $l=2$ in this subsection.

We can construct the generalized frame fields $E_A{}^I$ as given in Eq.~\eqref{eq:E-cH}. 
Then we can check that they satisfy the algebra $[E_A,\,E_B]_D = - \cF_{AB}{}^C\,E_C$ with the fluxes
\begin{align}
 \cF_{12}{}^3=-1\,,\qquad
 \cF_{13}{}^1=-1\,,\qquad
 \cF_{23}{}^2=1\,,\qquad
 \cF_{123} = -1\,.
\label{eq:SL2-WZW-alg}
\end{align}
We also obtain the constant metric $\hat{\cH}_{AB}$ by substituting Eq.~\eqref{eq:sl2-invg} into Eq.~\eqref{eq:E-cH}.
In the following, we perform the PL $T$-plurality by rotating the pair of the constants $\{\cF_{AB}{}^C,\,\hat{\cH}_{AB}\}$\,.

\subsubsection{Manin triple \texorpdfstring{$(\bm{4}|\bm{2.iii}|1)$}{(4|2.iii|1)}}

Let us consider an $\OO(3,3)$ transformation $T_A\to C_A{}^B\,T_B$ with
\begin{align}
 C_A{}^B = 
 {\footnotesize\begin{pmatrix}
 0 & 0 & 1 & 0 & 0 & 0 \\
 0 & 1 & 0 & 0 & 0 & 0 \\
 0 & 0 & 0 & 1 & 0 & 0 \\
 0 & 0 & 0 & 0 & 0 & 1 \\
 0 & 0 & 0 & 0 & 1 & 0 \\
 1 & 0 & 0 & 0 & 0 & 0\end{pmatrix} }.
\end{align}
Then we arrive at the Manin triple known as $(\bm{4}|\bm{2.iii}|b=1)$,
\begin{align}
 f_{12}{}^2 = -1\,,\qquad
 f_{12}{}^3 = 1\,,\qquad
 f_{13}{}^3 = -1\,,\qquad
 f_2{}^{13} = -1\,.
\end{align}
This shows that the flux algebra \eqref{eq:SL2-WZW-alg} corresponds to the Drinfel'd double, called DD7 \cite{math/0202210}, which admits six inequivalent Manin triples described in Eq.~\eqref{eq:DD7-orbit}. 

For $(\bm{4}|\bm{2.iii}|1)$, we parameterize the group element as
\begin{align}
 g= \Exp{x\,T_1}\Exp{y\,T_2+z\,T_3},
\end{align}
and then by using
\begin{align}
 r_m{}^a &= {\footnotesize\begin{pmatrix} 1 & 0 & 0 \\
 0 & \Exp{-x} & \Exp{-x} x \\
 0 & 0 & \Exp{-x}\end{pmatrix}},\qquad
 \ell_m{}^a = {\footnotesize\begin{pmatrix} 1 & -y & y-z \\
 0 & 1 & 0 \\
 0 & 0 & 1 \end{pmatrix}},
\\
 \pi^{ab} &= {\footnotesize\begin{pmatrix}
0 & 0 & \Exp{-x} y \\
 0 & 0 & \frac{y^2}{2} \Exp{-2 x} \\
 - \Exp{-x} y & -\frac{y^2}{2} \Exp{-2 x} & 0\end{pmatrix}},
\end{align}
the generalized metric and the DFT dilaton become
\begin{align}
 \cH_{IJ}(x) =
 {\footnotesize\begin{pmatrix}
 1 & 0 & 0 & 0 & 0 & y \\
 0 & 0 & 0 & 0 & -x & x^2-1 \\
 0 & 0 & 0 & 0 & -1 & x \\
 0 & 0 & 0 & 1 & y & -x y \\
 0 & -x & -1 & y & y^2 & -x y^2 \\
 y & x^2-1 & x & -x y & -x y^2 & x^2 y^2\end{pmatrix}},\qquad
 d(x)=0\,.
\end{align}
This (together with the spectator fields) satisfies the equations of motion of DFT. 

Again, this is a non-Riemannian background.
In order to get the standard description, let us we perform a factorized $T$-duality along the $z$-direction. 
Then we get
\begin{align}
 g_{mn} = {\footnotesize\begin{pmatrix}
 1 & 0 & y \\
 0 & 0 & -1 \\
 y & -1 & 0 \end{pmatrix}},\qquad
 B_{mn} = {\footnotesize\begin{pmatrix}
 0 & 0 & 0 \\
 0 & 0 & x \\
 0 & -x & 0 \end{pmatrix}},\qquad
 \Phi=0\,.
\end{align}
This background is (at least locally) the original AdS$_3$ background. 
Indeed, under a coordinate transformation
\begin{align}
 w \equiv -2\Exp{-x/2}\,,\quad
 x^+ \equiv -\Exp{-x}y\,,\quad
 x^- \equiv z\,, 
\end{align}
this three-dimensional background becomes
\begin{align}
 \rmd s^2 = 4\,\frac{\rmd w^2 + 2 \,\rmd x^+\,\rmd x^-}{w^2} \,,\qquad
 H_3 = \frac{8\,\rmd w\wedge \rmd x^+\wedge \rmd x^-}{w^3}\,.
\end{align}

The above result shows that the AdS$_3$ with $H$-flux has the $(\bm{4}|\bm{2.iii}|1)$ symmetry. 
One can explicitly construct the generators of the $(\bm{4}|\bm{2.iii}|1)$ symmetry as
\begin{align}
\begin{split}
 E_1 &= \partial_x\,,\qquad 
 E_2 = \Exp{x}\partial_y - \Exp{x}x\,\rmd z\,, \qquad
 E_3 = \Exp{x} \rmd z\,,
\\
 E^1 &= \rmd x + y\,\rmd z\,,\qquad
 E^2 = \Exp{-x}\rmd y + \frac{\Exp{-x}y^2}{2}\,\rmd z\,,
\\
 E^3 &= -\Exp{-x} y\,\partial_x -\frac{\Exp{-x} y^2}{2}\,\partial_y + \Exp{-x}\partial_z + \Exp{-x} x\,\rmd y + \frac{\Exp{-x} x y^2}{2}\,\rmd z\,.
\end{split}
\end{align}
We can easily check that they satisfy
\begin{align}
 [E_A,\,E_B]_D = - \cF_{AB}{}^C\,E_C\,, 
\end{align}
where $\cF_{AB}{}^C$ is the structure constant of the algebra $(\bm{4}|\bm{2.iii}|1)$. 

\subsubsection{Manin triple \texorpdfstring{$(\bm{2.iii}|\bm{4}|1)$}{(2.iii|4|1)}}

Here we consider the PL $T$-dual of the previous background. 
By using the parameterization
\begin{align}
 g = \Exp{z\,T_3} \Exp{(y+z)\,T_2} \Exp{x\,T_1},
\end{align}
and
\begin{align}
 r_m{}^a ={\footnotesize\begin{pmatrix}
 1 & z & 0 \\
 0 & 1 & 0 \\
 0 & 1 & 1
\end{pmatrix}},\qquad
 \ell_m{}^a={\footnotesize\begin{pmatrix}
 1 & 0 & 0 \\
 0 & 1 & 0 \\
 0 & x+1 & 1
\end{pmatrix}}, \qquad
 \pi^{ab} ={\footnotesize\begin{pmatrix}
 0 & y & z \\
 -y & 0 & \frac{z^2}{2} \\
 -z & -\frac{z^2}{2} & 0 \end{pmatrix}},
\end{align}
the dual background is found as
\begin{align}
 g_{mn} = {\footnotesize\begin{pmatrix}
 0 & 0 & \frac{1}{y} \\
 0 & \frac{1}{y^2} & \frac{1}{y^2} \\
 \frac{1}{y} & \frac{1}{y^2} & 0 \end{pmatrix}},\qquad
 B_{mn} = {\footnotesize\begin{pmatrix}
 0 & \frac{1}{y} & \frac{1}{y} \\
 -\frac{1}{y} & 0 & \frac{1}{y^2} \\
 -\frac{1}{y} & -\frac{1}{y^2} & 0 \end{pmatrix}},\qquad
 \Exp{-2\,\Phi}=y^2\,,\qquad
 I=\partial_x\,.
\end{align}
This satisfies the generalized supergravity equations of motion. 

Again we see that the three-dimensional geometry is AdS$_3$\,. 
Indeed, assuming $y>0$, a coordinate transformation
\begin{align}
 w=\sqrt{y}\,,\qquad x^+= x + \ln y\,,\qquad x^-=\tfrac{z}{4}\,,
\end{align}
and a $B$-field gauge transformation gives a ten-dimensional background
\begin{align}
\begin{split}
 \rmd s^2 &= 4\,\frac{\rmd w^2 + 2\,\rmd x^+\,\rmd x^-}{w^2} + \rmd s^2_{\text{S}^3\times T^4}\,,\qquad
 \Exp{-2\,\Phi}=w^4\,,
\\
 B_2 &= \frac{4\,\rmd x^+ \wedge\rmd x^-}{w^2} + \frac{2\,\rmd x^+\wedge\rmd w}{w} - \cos\theta\,\rmd\phi\wedge\rmd\psi\,,\qquad
 I = \partial_+ \,.
\end{split}
\label{eq:AdS-GSE}
\end{align}
This is precisely the solution obtained in \cite{1903.12175} [see Eq.~(4.11)] via the traditional non-Abelian $T$-duality (which is not based on the Drinfel'd double). 

\subsubsection{Manin triples \texorpdfstring{$(\bm{4}|\bm{7_0}|1)$ and $(\bm{4.i}|\bm{6_0}|-1)$}{(4|7\textzeroinferior|1) and (4.i|6\textzeroinferior|-1)}}

Here we consider an $\OO(3,3)$ transformation
\begin{align}
 C_A{}^B={\footnotesize\begin{pmatrix}
 0 & 0 & \sigma & 0 & 0 & 0 \\
 0 & 1 & 0 & 0 & 0 & 0 \\
 0 & 0 & 0 & 1 & 0 & 0 \\
 0 & 0 & 0 & 0 & 0 & \sigma \\
 0 & 0 & 0 & - \frac{\sigma}{2} & 1 & 0 \\
 1 & \frac{\sigma}{2} & 0 & 0 & 0 & 0\end{pmatrix}},\qquad \sigma=\pm 1\,,
\end{align}
of the algebra \eqref{eq:SL2-WZW-alg}.
Then we arrive at the algebra
\begin{align}
 \cF_{12}{}^2 = -\sigma\,,\quad
 \cF_{12}{}^3 = \sigma\,,\quad
 \cF_{13}{}^3 = -\sigma\,,\quad
 \cF_1{}^{23} = 1\,,\quad
 \cF_2{}^{13} = -\sigma\,.
\end{align}
This corresponds to the Manin triple $(\bm{4}|\bm{7_0}|1)$ or $(\bm{4.i}|\bm{6_0}|-1)$ for $\sigma=+1$ or $-1$\,, respectively. 
The constant matrix becomes
\begin{align}
 \hat{\cH}_{AB}=
{\footnotesize\begin{pmatrix}
 1 & 0 & 0 & 0 & 0 & 0 \\
 0 & 0 & 0 & 0 & 0 & -1 \\
 0 & 0 & 0 & 0 & -1 & 0 \\
 0 & 0 & 0 & 1 & 0 & 0 \\
 0 & 0 & -1 & 0 & \sigma & 0 \\
 0 & -1 & 0 & 0 & 0 & -\sigma \end{pmatrix}}.
\end{align}
By using
\begin{align}
 g= \Exp{x\,T_1}\Exp{y\,T_2 + z\,T_3},
\end{align}
we obtain
\begin{align}
 r_m{}^a&={\footnotesize\begin{pmatrix}
 1 & 0 & 0 \\
 0 & \Exp{-\sigma x} & \sigma\,x \Exp{-\sigma x} \\
 0 & 0 & \Exp{-\sigma x} 
\end{pmatrix}},\qquad
 \ell_m{}^a={\footnotesize\begin{pmatrix}
 1 & y & - y+z \\
 0 & 1 & 0 \\
 0 & 0 & 1 
\end{pmatrix}},
\\
 \pi^{ab}&={\footnotesize\begin{pmatrix}
 0 & 0 & \sigma\,y\Exp{-\sigma x} \\
 0 & 0 & \frac{\Exp{-2\sigma x} (\sigma+y^2)-\sigma}{2} \\
 -\sigma\,y\Exp{-\sigma\,x} & -\frac{\Exp{-2\sigma x} (\sigma+y^2)-\sigma}{2} & 0 \end{pmatrix}}.
\end{align}
Then the supergravity fields are
\begin{align}
 g_{mn}={\footnotesize\begin{pmatrix}
 1+\sigma\,y^2 & -y & 0 \\
 - y & \sigma\,(1-x^2) & - x \\
 0 & - x & -\sigma \end{pmatrix}},\qquad
 B_{mn}={\footnotesize\begin{pmatrix}
 0 & -\sigma\, x\,y & -y \\
 \sigma\,x\,y & 0 & \sigma \\
 y & -\sigma & 0 
\end{pmatrix}}, \qquad \Phi=0\,.
\end{align}
By computing the curvature tensor and the $H$-flux, we can identify that this is again the original AdS$_3$ background with the AdS radius $l=2$\,. 

\subsubsection{Manin triples \texorpdfstring{$(\bm{7_0}|\bm{4}|1)$ and $(\bm{6_0}|\bm{4.i}|-1)$}{(7\textzeroinferior|4|1) and (6\textzeroinferior|4.i|-1)}}

Let us also consider the PL $T$-dual of the previous example, which corresponds to the Manin triple $(\bm{7_0}|\bm{4}|1)$ or $(\bm{6_0}|\bm{4.i}|-1)$ for $\sigma=+1$ or $-1$\,, respectively. 
Using a parameterization,
\begin{align}
 g= \Exp{y\,T_3}\Exp{x\,T_1 + (y-z)\,T_2},
\end{align}
we find
\begin{align}
 r_m{}^a&={\footnotesize\begin{pmatrix}
 \cos y & \sin y & 0 \\
 -\sin y & \cos y & 1 \\
 0 & 0 & 1 
\end{pmatrix}},\qquad
 \ell_m{}^a={\footnotesize\begin{pmatrix}
 1 & 0 & 0 \\
 z-y & x+1 & 1 \\
 z-y & x & 1 
\end{pmatrix}},
\\
 \pi^{ab}&={\footnotesize\begin{pmatrix}
 0 & -z & \sin y \\
 z & 0 & 1-\cos y \\
 -\sin y & \cos y -1 & 0 \end{pmatrix}} ,
\end{align}
for $\sigma=+1$ and
\begin{align}
 r_m{}^a&={\footnotesize\begin{pmatrix}
 \cosh y & -\sinh y & 0 \\
 -\sinh y & \cosh y & 1 \\
 0 & 0 & 1 \end{pmatrix}},\qquad
 \ell_m{}^a={\footnotesize\begin{pmatrix}
 1 & 0 & 0 \\
 z-y & 1-x & 1 \\
 z-y & -x & 1 \end{pmatrix}},
\\
 \pi^{ab}&={\footnotesize\begin{pmatrix}
 0 & z & -\sinh y \\
 -z & 0 & \cosh y -1 \\
 \sinh y & 1-\cosh y & 0 \end{pmatrix}} ,
\end{align}
for $\sigma=-1$\,.
Considering that the dual structure constants have non-vanishing trace, we find a solution of the generalized supergravity equations of motion
\begin{align}
\begin{split}
 g_{mn}&=\begin{pmatrix}
 0 & -\frac{\sigma }{z} & -\frac{\sigma }{z} \\
 -\frac{\sigma }{z} & -\sigma & -\sigma -\frac{1}{z^2} \\
 -\frac{\sigma }{z} & -\sigma -\frac{1}{z^2} & -\sigma -\frac{1}{z^2} 
\end{pmatrix},\qquad
 B_{mn}=\begin{pmatrix}
 0 & -\frac{\sigma}{z} & 0 \\
 \frac{\sigma}{z} & 0 & \frac{1}{z^2} \\
 0 & -\frac{1}{z^2} & 0 
\end{pmatrix},
\\
 \Exp{-2\,\Phi} &= z^2\,,\qquad I = \sigma\,\partial_x\,.
\end{split}
\end{align}
If we perform a coordinate transformation,
\begin{align}
 x \equiv \sigma\,(x^+ - 2 \ln w)\,,\qquad y\equiv -4\,x^- - w^2\,, \qquad z\equiv w^2\,,
\end{align}
we obtain
\begin{align}
\begin{split}
 \rmd s^2 &= 4\,\frac{\rmd w^2 + 2\,\rmd x^+\,\rmd x^-}{w^2} - 16\,\sigma\,(\rmd x^-)^2 \,,\qquad
 \Exp{-2\,\Phi}=w^4\,,
\\
 B_2 &= \frac{4\,\rmd x^+ \wedge\rmd x^-}{w^2} + \frac{2\,\rmd x^+\wedge\rmd w}{w} \,,\qquad
 I = \partial_+ \,.
\end{split}
\end{align}
This AdS$_3$ solution will be the same as the solution given in Eq.~\eqref{eq:AdS-GSE} up to a further coordinate transformation. 

\section{Nappi--Witten model}
\label{sec:NW}

The NW model \cite{hep-th:9310112} is the WZW model based on a central extension of the two-dimensional Euclidean group $E_2^c$
\begin{align}
 [J,\,P_i]=\epsilon_{ij}\,P_j\,,\qquad 
 [P_i,\,P_j]=\epsilon_{ij}\,T\,.
\end{align}
We denote the generators collectively as
\begin{align}
 \{T_a\} = \{P_1,\,P_2,\,J,\,T\}\,,
\end{align}
and parameterize the group element as \cite{hep-th:9310112}
\begin{align}
 g=\Exp{x\,T_1+y\,T_2}\Exp{u\,T_3+v\,T_4}.
\label{eq:g-NW-param}
\end{align}
The right- and left-invariant 1-forms are
\begin{align}
 r_m{}^a = {\footnotesize\begin{pmatrix}
 1 & 0 & 0 & -\frac{y}{2} \\
 0 & 1 & 0 & \frac{x}{2} \\
 y & -x & 1 & -\frac{x^2+y^2}{2} \\
 0 & 0 & 0 & 1 
\end{pmatrix}},\qquad
 \ell_m{}^a = {\footnotesize\begin{pmatrix}
 \cos u & -\sin u & 0 & \frac{y}{2} \\
 \sin u & \cos u & 0 & -\frac{x}{2} \\
 0 & 0 & 1 & 0 \\
 0 & 0 & 0 & 1 \end{pmatrix}}.
\label{eq:NW-rl}
\end{align}
Using the non-degenerate invariant metric
\begin{align}
 \hat{g}_{ab} = {\footnotesize\begin{pmatrix}
 1 & 0 & 0 & 0 \\
 0 & 1 & 0 & 0 \\
 0 & 0 & b & 1 \\
 0 & 0 & 1 & 0
\end{pmatrix}},
\label{eq:NW-ghat}
\end{align}
we obtain the NW background
\begin{align}
 g_{mn}={\footnotesize \begin{pmatrix}
 1 & 0 & \frac{y}{2} & 0 \\
 0 & 1 & -\frac{x}{2} & 0 \\
 \frac{y}{2} & -\frac{x}{2} & b & 1 \\
 0 & 0 & 1 & 0 \end{pmatrix}}, \qquad 
 H_3 = \frac{1}{3!}\,f_{abc}\,r^a\wedge r^b\wedge r^c 
 = \rmd x\wedge \rmd y\wedge \rmd u\,.
\label{eq:NW}
\end{align}
Choosing the $B$-field, for example, as
\begin{align}
 B_2 = u\,\rmd x\wedge \rmd y\,,
\end{align}
we can construct the generalized frame fields $E_A{}^I$ \eqref{eq:E-cH} satisfying the algebra \eqref{eq:gen-Lie-algebra} with
\begin{align}
 \cF_{12}{}^4 = 1\,,\quad 
 \cF_{13}{}^2 = -1\,,\quad 
 \cF_{23}{}^1 = 1\,,\quad 
 \cF_{123} = 1\,.
\label{eq:NW-alg}
\end{align}
Using the constant metric $\hat{\cH}_{AB}$\,, defined by Eq.~\eqref{eq:E-cH}, we obtain the generalized metric
\begin{align}
 \cH_{IJ} = E_I{}^A\,E_J{}^B\,\hat{\cH}_{AB}\,,
\end{align}
which describes the NW background \eqref{eq:NW}. 

The fluxes $\cF_{AB}{}^C$ and $\hat{\cH}_{AB}$ satisfy the equations of motion \eqref{eq:eom}, and in the following subsections, we consider several $\OO(4,4)$ rotations $T_A\to C_A{}^B\,T_B$ and $\hat{\cH}_{AB}\to C_A{}^C\,C_B{}^D\,\hat{\cH}_{CD}$ to find the dual solutions. 

\subsection{Semi-Abelian double \texorpdfstring{$(E_2^c|A_4)$}{(E\texttwoinferior|A\textfourinferior)}}

As the first example, let us consider an $\OO(4,4)$ transformation
\begin{align}
 C_A{}^B = {\footnotesize\begin{pmatrix}
 1 & 0 & 0 & 0 & 0 & 0 & 0 & 0 \\
 0 & 1 & 0 & 0 & 0 & 0 & 0 & 0 \\
 0 & 0 & 1 & 0 & 0 & 0 & 0 & -1 \\
 0 & 0 & 0 & 1 & 0 & 0 & 1 & 0 \\
 0 & 0 & 0 & 0 & 1 & 0 & 0 & 0 \\
 0 & 0 & 0 & 0 & 0 & 1 & 0 & 0 \\
 0 & 0 & 0 & 0 & 0 & 0 & 1 & 0 \\
 0 & 0 & 0 & 0 & 0 & 0 & 0 & 1 
\end{pmatrix}}.
\end{align}
The original fluxes \eqref{eq:NW-alg} are mapped to the fluxes
\begin{align}
 \cF_{12}{}^4 = 1\,,\quad 
 \cF_{13}{}^2 = -1\,,\quad 
 \cF_{23}{}^1 = 1\,.
\label{eq:NW-alg-H0}
\end{align}
The (geometric) fluxes $\cF_{ab}{}^c$ are precisely the original ones and the $H$-flux disappeared under the $\OO(4,4)$ transformation. 
This is an algebra of the Drinfel'd double (especially a semi-Abelian double $\cF_a{}^{bc}=0$) and we denote the Manin triple as $(E_2^c|A_4)$\,, where $A_4$ denotes the four-dimensional Abelian algebra. 
We can easily construct the generalized frame fields by using the group element \eqref{eq:g-NW-param}. 
In this frame, the constant metric becomes
\begin{align}
 \hat{\cH}_{AB}=
 {\footnotesize\begin{pmatrix}
 1 & 0 & 0 & 0 & 0 & 0 & 0 & 0 \\
 0 & 1 & 0 & 0 & 0 & 0 & 0 & 0 \\
 0 & 0 & 0 & 0 & 0 & 0 & -1 & b \\
 0 & 0 & 0 & 0 & 0 & 0 & 0 & 1 \\
 0 & 0 & 0 & 0 & 1 & 0 & 0 & 0 \\
 0 & 0 & 0 & 0 & 0 & 1 & 0 & 0 \\
 0 & 0 & -1 & 0 & 0 & 0 & 0 & 1 \\
 0 & 0 & b & 1 & 0 & 0 & 1 & -b \end{pmatrix}}.
\label{eq:NW-Hhat}
\end{align}
Then we find the supergravity fields as
\begin{align}
 g_{mn}={\footnotesize \begin{pmatrix}
 1 & 0 & \frac{y}{2} & 0 \\
 0 & 1 & -\frac{x}{2} & 0 \\
 \frac{y}{2} & -\frac{x}{2} & b & 1 \\
 0 & 0 & 1 & 0 \end{pmatrix}},\qquad
 B_{mn}= {\footnotesize \begin{pmatrix}
 0 & 0 & -\frac{y}{2} & 0 \\
 0 & 0 & \frac{x}{2} & 0 \\
 \frac{y}{2} & -\frac{x}{2} & 0 & -1 \\
 0 & 0 & 1 & 0 
\end{pmatrix}},\qquad \Phi=0\,.
\label{eq:NW-2}
\end{align}
The $H$-flux is
\begin{align}
 H_3 = \rmd x\wedge\rmd y\wedge \rmd u\,,
\end{align}
and it turns out that this background is precisely the original NW background. 
Here, the condition $\Lie_{v_a}E_{mn}=0$ for the PL symmetry is satisfied, and we can perform the PL $T$-duality/plurality as usual. 

\subsection{Semi-Abelian double \texorpdfstring{$(A_4|E_2^c)$}{(A\textfourinferior|E\texttwoinferior)}}

Here we consider the PL $T$-duality of the previous example, where the Manin triple can be denoted as $(A_4|E_2^c)$. 
By using the parameterization $g=\Exp{x\,T_1+y\,T_2+u\,T_3+v\,T_4}$\,, we find
\begin{align}
 r_m^a=\ell_m^a=\delta_m^a\,,\qquad 
 \pi^{ab} = {\footnotesize\begin{pmatrix}
  0 & v & -y & 0 \\
  -v & 0 & x & 0 \\
  y & -x & 0 & 0 \\
  0 & 0 & 0& 0 \end{pmatrix}} .
\end{align}
The generalized metric cannot be parameterized by $g_{mn}$ and $B_{mn}$\,, but we find a solution of DFT,
\begin{align}
 \cH_{IJ}(x) = \begin{pmatrix} G_{mn} & G_{mp}\,\beta^{pn} \\ -\beta^{mp}\,G_{pn} & G^{mn}-\beta^{mp}\,G_{pq}\,\beta^{qn}\end{pmatrix} ,\qquad d(x)=0\,,
\end{align}
where
\begin{align}
 G_{mn}(x)={\footnotesize\begin{pmatrix}
 1 & 0 & 0 & 0 \\
 0 & 1 & 0 & 0 \\
 0 & 0 & 0 & 1 \\
 0 & 0 & 1 & -b \end{pmatrix}},\qquad
 \beta^{mn}(x)={\footnotesize\begin{pmatrix}
 0 & -v & y & 0 \\
 v & 0 & -x & 0 \\
 -y & x & 0 & 1 \\
 0 & 0 & -1 & 0 \end{pmatrix}}.
\label{eq:nR-NW1}
\end{align}
The open-string metric is flat and there is the constant $Q$-flux
\begin{align}
 Q_x{}^{yu}=-1\,,\qquad Q_y{}^{xu}=1\,,\qquad Q_v{}^{xy}=-1\,. 
\end{align}
If we make a periodic identification, such as $x\sim x+1$\,, this spacetime can be regarded a $T$-fold \cite{hep-th:0406102}. 
In this example, it is not easy to find an Abelian $\OO(4,4)$ transformation which brings this non-Riemannian background into a Riemannian one.

\subsection{Manin triple $(G1|G2)$}

From the algebra \eqref{eq:NW-alg-H0}, by performing an $\OO(4,4)$ transformation,
\begin{align}
 C_A{}^B = {\footnotesize\begin{pmatrix}
 1 & 0 & 0 & 0 & 0 & 0 & 0 & 0 \\
 0 & 1 & 0 & 0 & 0 & 0 & 0 & 0 \\
 0 & 0 & 0 & 0 & 0 & 0 & 1 & 0 \\
 0 & 0 & 0 & 1 & 0 & 0 & 0 & 0 \\
 0 & 0 & 0 & 0 & 1 & 0 & 0 & 0 \\
 0 & 0 & 0 & 0 & 0 & 1 & 0 & 0 \\
 0 & 0 & 1 & 0 & 0 & 0 & 0 & 0 \\
 0 & 0 & 0 & 0 & 0 & 0 & 0 & 1 \end{pmatrix}},
\end{align}
we obtain another Manin triple
\begin{align}
 \cF_{12}{}^4 = 1\,,\quad
 \cF_2{}^{13} = -1\,,\quad
 \cF_1{}^{23} = 1\,.
\end{align}
For convenience, we denote the four-dimensional algebras, characterized by $\cF_{ab}{}^c$ and $\cF_c{}^{ab}$\,, by $G1$ and $G2$\,, respectively. 
The constant matrix becomes
\begin{align}
 \hat{\cH}_{AB} = {\footnotesize\begin{pmatrix}
 1 & 0 & 0 & 0 & 0 & 0 & 0 & 0 \\
 0 & 1 & 0 & 0 & 0 & 0 & 0 & 0 \\
 0 & 0 & 0 & 0 & 0 & 0 & -1 & 1 \\
 0 & 0 & 0 & 0 & 0 & 0 & 0 & 1 \\
 0 & 0 & 0 & 0 & 1 & 0 & 0 & 0 \\
 0 & 0 & 0 & 0 & 0 & 1 & 0 & 0 \\
 0 & 0 & -1 & 0 & 0 & 0 & 0 & b \\
 0 & 0 & 1 & 1 & 0 & 0 & b & -b \end{pmatrix}}\,.
\end{align}
As one can easily expect, we get a non-Riemannian background when $b=0$\,. 
Assuming $b\neq 0$ and using a parameterization $g=\Exp{x\,T_1+y\,T_2 -(b\,u+2\,v)\,T_3 + v\,T_4}$\,, we find
\begin{align}
 r_m{}^a&={\footnotesize\begin{pmatrix}
 1 & 0 & 0 & -\frac{y}{2} \\
 0 & 1 & 0 & \frac{x}{2} \\
 0 & 0 & -b & 0 \\
 0 & 0 & -2 & 1 \end{pmatrix}},\qquad
 \ell_m{}^a={\footnotesize\begin{pmatrix}
 1 & 0 & 0 & \frac{y}{2} \\
 0 & 1 & 0 & -\frac{x}{2} \\
 0 & 0 & -b & 0 \\
 0 & 0 & -2 & 1 \end{pmatrix}},
\\
 \pi^{ab}&={\footnotesize\begin{pmatrix}
 0 & 0 & y & 0 \\
 0 & 0 & -x & 0 \\
 -y & x & 0 & \frac{x^2+y^2}{2} \\
 0 & 0 & -\frac{x^2+y^2}{2} & 0 \end{pmatrix}},
\end{align}
and the dual background is precisely the original one
\begin{align}
 g_{mn} = {\footnotesize\begin{pmatrix}
 1 & 0 & \frac{y}{2} & 0 \\
 0 & 1 & -\frac{x}{2} & 0 \\
 \frac{y}{2} & -\frac{x}{2} & b & 1 \\
 0 & 0 & 1 & 0 \end{pmatrix}},\qquad
 B_{mn} = {\footnotesize\begin{pmatrix}
 0 & 0 & -\frac{y}{2} & 0 \\
 0 & 0 & \frac{x}{2} & 0 \\
 \frac{y}{2} & -\frac{x}{2} & 0 & 1 \\
 0 & 0 & -1 & 0 \end{pmatrix}},\qquad \Exp{-2\,\Phi}= b^{-1}\,,
\end{align}
up to a $B$-field gauge transformation and a constant shift of the dilaton. 
Then, we find the invariance of the NW background under the PL $T$-plurality. 

\subsection{Manin triple $(G2|G1)$}

The PL $T$-duality of the previous example gives a Manin triple $(G2|G1)$\,. 
There, again assuming $b\neq 0$ and using $g=\Exp{\frac{u}{b}\,T_3}\Exp{x\,T_1+y\,T_2+v\,T_4}$, we obtain
\begin{align}
 r_m{}^a ={\footnotesize\begin{pmatrix}
 \cos \frac{u}{b} & \sin \frac{u}{b} & 0 & 0 \\
 -\sin \frac{u}{b} & \cos \frac{u}{b} & 0 & 0 \\
 0 & 0 & \frac{1}{b} & 0 \\
 0 & 0 & 0 & 1 \end{pmatrix}},\quad
 \ell_m{}^a={\footnotesize\begin{pmatrix}
 1 & 0 & 0 & 0 \\
 0 & 1 & 0 & 0 \\
 -\frac{y}{b} & \frac{x}{b} & \frac{1}{b} & 0 \\
 0 & 0 & 0 & 1 \end{pmatrix}},\quad
 \pi^{ab}={\footnotesize\begin{pmatrix}
 0 & -v & 0 & 0 \\
 v & 0 & 0 & 0 \\
 0 & 0 & 0 & 0 \\
 0 & 0 & 0 & 0 \end{pmatrix}}.
\end{align}
The dual geometry is non-Riemannian and we find
\begin{align}
 G_{mn}(x)={\footnotesize\begin{pmatrix}
 1 & 0 & 0 & 0 \\
 0 & 1 & 0 & 0 \\
 0 & 0 & 0 & 1 \\
 0 & 0 & 1 & -b \end{pmatrix}},\qquad
 \beta^{mn}(x)={\footnotesize\begin{pmatrix}
 0 & -v & 0 & 0 \\
 v & 0 & 0 & 0 \\
 0 & 0 & 0 & 1 \\
 0 & 0 & -1 & 0 \end{pmatrix}},\qquad
 \Exp{-2\,d(x)}=b^{-1}\,.
\end{align}
In fact, we find that
\begin{align}
 G_{mn}(x)={\footnotesize\begin{pmatrix}
 1 & 0 & 0 & 0 \\
 0 & 1 & 0 & 0 \\
 0 & 0 & 0 & 1 \\
 0 & 0 & 1 & -b \end{pmatrix}},\qquad
 \beta^{mn}(x)={\footnotesize\begin{pmatrix}
 0 & -v & -c_0\,y & 0 \\
 v & 0 & c_0\,x & 0 \\
 c_0\,y & -c_0\,x & 0 & 1 \\
 0 & 0 & -1 & 0 \end{pmatrix}},\qquad
 \Exp{-2\,d(x)}=\text{const.}\,,
\label{eq:nR-NW2}
\end{align}
solves the equations of motion for an arbitrary parameter $c_0$\,. 
Both \eqref{eq:nR-NW1} and \eqref{eq:nR-NW2} are contained in this one-parameter family of solutions.

\subsection{Yang--Baxter deformation based on modified CYBE}

Here we consider the YB deformation of the NW background \eqref{eq:NW-2}. 
YB deformations of the NW model were studied in \cite{1511.00404} by following the prescription of \cite{1410.8066}. 
There, the general solution of the (modified) CYBE was found, but the deformation can be removed by a coordinate transformation and a $B$-field gauge transformation, and a new background was not found. 
In other words, the NW background was found to be invariant (or self-dual) under the YB deformation. 

When the $r$-matrix solves the homogeneous CYBE, the YB deformation is a particular case of the PL $T$-plurality. 
In the case of the NW background, the general solution of the homogeneous CYBE is Abelian \cite{1511.00404} and then the deformation is just an Abelian $T$-duality transformation. 
The only non-Abelian solution can be found by considering a solution of the modified CYBE. 
Here, we consider the YB deformation by using a solution of the modified CYBE. 
It seems that our deformation is different from the one studied in \cite{1511.00404}, and we find a solution which connects the NW background and a flat solution. 

We consider a Lie algebra of the Drinfel'd double where the dual structure constants satisfy the coboundary ansatz,
\begin{align}
 f_a{}^{bc} = f_{ad}{}^b\,r^{dc} - f_{ad}{}^c\,r^{db}\qquad \bigl(r^{ab}=-r^{ba}\bigr)\,.
\end{align}
This Drinfel'd double satisfies the Jacobi identity if the following (modified) CYBE is satisfied:
\begin{align}
 f_{d_1d_2}{}^{a_1}\,r^{d_1a_2}\,r^{d_2a_3}
 +f_{d_1d_2}{}^{a_2}\,r^{d_1a_3}\,r^{d_2a_1}
 +f_{d_1d_2}{}^{a_3}\,r^{d_1a_1}\,r^{d_2a_2}
 = c^2\,f_{d_1d_2}{}^{a_1}\,\hat{g}^{d_1a_2}\,\hat{g}^{d_2a_3}\,,
\label{eq:mCYBE}
\end{align}
where $\hat{g}^{ab}$ is the inverse matrix of an invariant metric $\hat{g}_{ab}$ satisfying $f_{ca}{}^d\,\hat{g}_{db}+f_{cb}{}^d\,\hat{g}_{ad}=0$\,. 
Here, we consider the Lie algebra $f_{ab}{}^c$ given in Eq.~\eqref{eq:NW-alg-H0} and the invariant metric
\begin{align}
 \hat{g}_{ab} = {\footnotesize\begin{pmatrix} 1 & 0 & 0 & 0 \\ 0 & 1 & 0 & 0 \\ 0 & 0 & b & 1 \\ 0 & 0 & 1 & 0 \end{pmatrix}}.
\end{align}
The general solution of \eqref{eq:mCYBE} was found in \cite{1511.00404}, and the only non-Abelian solution is
\begin{align}
 r^{12} = \frac{c}{2}\,,
\label{eq:r-mat}
\end{align}
which gives the structure constants
\begin{align}
 f_{12}{}^4 = 1\,,\quad 
 f_{13}{}^2 = -1\,,\quad 
 f_{23}{}^1 = 1\,,\quad
 f_1{}^{14} = \eta\,,\quad
 f_2{}^{24} = \eta \qquad (\eta\equiv c/2)\,.
\label{eq:f-eta}
\end{align}

The YB deformation can be understood as the deformation of the algebra from $\eta=0$ to $\eta=c/2$\,.
When the $r$-matrix satisfies the homogeneous CYBE ($c=0$), a YB deformation is precisely an $\OO(4,4)$ transformation
\begin{align}
 T_A \to C_A{}^B\,T_B\,,\qquad C_A{}^B = \begin{pmatrix} \delta_a^b & 0 \\ r^{ab} & \delta^a_b \end{pmatrix}.
\label{eq:O44-eta}
\end{align}
However, when the $r$-matrix satisfies the modified CYBE, it is not an $\OO(4,4)$ transformation. 
Indeed, if we perform the inverse transformation of \eqref{eq:O44-eta}, the algebra \eqref{eq:f-eta} does not go back to the original one \eqref{eq:NW-alg-H0}. 
Rather, the algebra becomes
\begin{align}
 \cF_{12}{}^4 = 1\,,\quad 
 \cF_{13}{}^2 = -1\,,\quad 
 \cF_{23}{}^1 = 1\,,\quad
 \cF^{124} = -\eta^2 \,,
\end{align}
and this includes the $R$-flux $\cF^{abc}\,(=\cF^{[abc]})$ in addition to the original geometric flux (see, for example, \cite{1803.03971} for more details of the generalized fluxes).

Fortunately, our constant metric \eqref{eq:NW-Hhat},
\begin{align}
 \hat{\cH}_{AB}=
 {\footnotesize\begin{pmatrix}
 1 & 0 & 0 & 0 & 0 & 0 & 0 & 0 \\
 0 & 1 & 0 & 0 & 0 & 0 & 0 & 0 \\
 0 & 0 & 0 & 0 & 0 & 0 & -1 & b \\
 0 & 0 & 0 & 0 & 0 & 0 & 0 & 1 \\
 0 & 0 & 0 & 0 & 1 & 0 & 0 & 0 \\
 0 & 0 & 0 & 0 & 0 & 1 & 0 & 0 \\
 0 & 0 & -1 & 0 & 0 & 0 & 0 & 1 \\
 0 & 0 & b & 1 & 0 & 0 & 1 & -b \end{pmatrix}},
\end{align}
satisfies the equations of motion of DFT \eqref{eq:eom} for an arbitrary value of the $R$-flux $\cF^{124}$\,. 
Thus, the fluxes \eqref{eq:f-eta} together with the $\OO(4,4)$-rotated constant matrix,
\begin{align}
 \hat{\cH}_{AB}=
 {\footnotesize\begin{pmatrix} 1 & 0 & 0 & 0 & 0 & -\eta & 0 & 0 \\
 0 & 1 & 0 & 0 & \eta & 0 & 0 & 0 \\
 0 & 0 & 0 & 0 & 0 & 0 & -1 & b \\
 0 & 0 & 0 & 0 & 0 & 0 & 0 & 1 \\
 0 & \eta & 0 & 0 & \eta^2+1 & 0 & 0 & 0 \\
 -\eta & 0 & 0 & 0 & 0 & \eta^2+1 & 0 & 0 \\
 0 & 0 & -1 & 0 & 0 & 0 & 0 & 1 \\
 0 & 0 & b & 1 & 0 & 0 & 1 & -b \end{pmatrix}},
\end{align}
satisfies the equations of motion of DFT. 

Now let us explicitly construct the YB-deformed background. 
Using the group element
\begin{align}
 g=\Exp{x\,T_1+y\,T_2}\Exp{u\,T_3+v\,T_4},
\end{align}
$r_m^a$ and $\ell_m^a$ are obtained as in Eq.~\eqref{eq:NW-rl} and we also find
\begin{align}
 \pi^{ab} = {\footnotesize\begin{pmatrix}
 0 & 0 & 0 & -\eta\,x \\
 0 & 0 & 0 & -\eta\,y \\
 0 & 0 & 0 & 0 \\
 \eta\,x & \eta\,y & 0 & 0 \end{pmatrix}} .
\end{align}
Since the dual structure constants are non-unimodular $f_b{}^{ba}\neq 0$, we find a solution of the generalized supergravity equations of motion
\begin{align}
\begin{split}
 g_{mn} &= {\footnotesize\begin{pmatrix}
 \frac{1}{1+\eta^2} & 0 & \frac{y}{2}-\frac{\eta\,x}{1+\eta^2} & 0 \\
 0 & \frac{1}{1+\eta^2} & -\frac{x}{2}-\frac{\eta y}{1+\eta^2} & 0 \\
 \frac{y}{2}-\frac{\eta x}{1+\eta^2} & -\frac{x}{2}-\frac{\eta y}{1+\eta^2} & b+\frac{\eta^2\,(x^2+y^2)}{\eta^2+1} & 1 \\
 0 & 0 & 1 & 0 \end{pmatrix}},\qquad \Exp{-2\,\Phi}= 1 + \eta^2\,,
\\
 B_{mn} &= {\footnotesize\begin{pmatrix}
 0 & -\frac{\eta}{1+\eta^2} & \frac{y}{2}-\frac{y}{\eta^2+1} & 0 \\
 \frac{\eta}{1+\eta^2} & 0 & \frac{x}{1+\eta^2}-\frac{x}{2} & 0 \\
 \frac{y}{1+\eta^2}-\frac{y}{2} & \frac{x}{2}-\frac{x}{1+\eta^2} & 0 & -1 \\
 0 & 0 & 1 & 0 \end{pmatrix}},\qquad I= \tfrac{1}{2}\,f_b{}^{ba}\,v_a = \eta\,v_4 = \eta\,\partial_v\,.
\end{split}
\label{eq:eta-sol}
\end{align}
Interestingly, due to the degeneracy $(g+B)_{mn}\,I^n=0$\,, this vector field $I$ disappears from the equations of motion and can be removed.\footnote{A similar situation has been observed in \cite{1803.05903} and studied in more detail in \cite{1803.07391}.} 
Consequently we obtain a one-parameter family of supergravity solutions \eqref{eq:eta-sol} without $I$\,. 

By the construction, this solution reduces to the NW background by choosing $\eta=0$\,. 
Moreover, we find an interesting special case $\eta=\pm 1$ 
\begin{align}
\begin{split}
 g_{mn} &= {\footnotesize\begin{pmatrix}
 \frac{1}{2} & 0 & \frac{y-\eta\,x}{2} & 0 \\
 0 & \frac{1}{2} & -\frac{x+\eta\,y}{2} & 0 \\
 \frac{y-\eta\,x}{2} & -\frac{x+\eta\,y}{2} & b+\frac{x^2+y^2}{2} & 1 \\
 0 & 0 & 1 & 0\end{pmatrix}}, \quad
 B_{mn} = {\footnotesize\begin{pmatrix}
 0 & -\frac{\eta}{2} & 0 & 0 \\
 \frac{\eta}{2} & 0 & 0 & 0 \\
 0 & 0 & 0 & -1 \\
 0 & 0 & 1 & 0 \end{pmatrix}},\quad
 \Exp{-2\,\Phi}= 2\,,
\end{split}
\end{align}
where the curvature tensor and the $H$-flux vanish.
Therefore, we found a one-parameter family of solutions which contains the NW background ($\eta=0$) and the four-dimensional Minkowski spacetime ($\eta=\pm1$) as specific cases. 

\section{Conclusions}
\label{sec:conclusion}

The main purpose of this paper is to point out that the target space of a WZW model can be used as a seed solution to generate a chain of solutions through the PL $T$-plurality.
The generalized frame fields $E_A{}^I$ and the constant metric $\hat{\cH}_{AB}$ given in Eq.~\eqref{eq:E-cH} construct the WZW background, and additionally $E_A{}^I$ satisfy the algebra $[E_A,\,E_B]_D=-\cF_{AB}{}^C\,E_C$ associated with the $2D$-dimensional algebra
\begin{align}
 \cF_{abc}=H_{abc}\,,\qquad \cF_{ab}{}^c=\cF_b{}^c{}_a=\cF^c{}_{ab}=f_{ab}{}^c\,,\qquad
 \cF_a{}^{bc} = \cF^c{}_a{}^b=\cF^{bc}{}_a=0=\cF^{abc}\,.
\end{align}
If we find an $\OO(D,D)$ transformation which transforms this algebra into a Drinfel'd double, we obtain a PL symmetric background, and we can construct further PL symmetric backgrounds by following the standard procedure of the PL $T$-plurality. 
As demonstrations, we studied the PL $T$-plurality for two WZW backgrounds, AdS$_3$ with $H$-flux and the NW background. 

In the case of the AdS$_3$ with $H$-flux, we considered two Drinfel'd doubles DD2 and DD7. 
There are $4+6$ inequivalent Manin triples, but all of them are related to the following four solutions through a coordinate transformation or the standard Abelian $T$-duality:
\begin{align}
 \text{Sol1:}\qquad & \rmd s^2 = l^2\,\frac{\rmd z^2 - \rmd t^2 + \rmd x^2}{z^2}\,, \quad H_3 = \pm 2\,l^{-1} *1 \,,\quad \Phi=0\,,
\\
 \text{Sol2:}\qquad & \rmd s^2_{\text{(open)}} = \rmd x^2 +4\,\bigl(\rmd y + y\,\rmd x\bigr)\,\rmd z\,,\quad
 \beta = \bigl(y\,\partial_x-\tfrac{1}{2}\,\partial_z\bigr)\wedge\partial_y\,,\quad d(x)=-x\,,
\\
 \text{Sol3:}\qquad & \rmd s^2 = 2\,\rmd x\,\rmd y+\rmd z^2 \,,\quad B_2 =0\,,\quad \Phi = \pm z\,,
\\
 \text{Sol4:}\qquad & \rmd s^2 = l^2\,\frac{\rmd z^2 + 2\,\rmd x^+\,\rmd x^-}{z^2}+\lambda\,(\rmd x^-)^2\,,\quad
 B_2 = \frac{\rmd x^+ \wedge (l^2\,\rmd x^- + 2\,z\,\rmd z)}{z^2} \,,
\nonumber\\
 &\Exp{-2\,\Phi}=z^4\,,\quad I = \partial_+ \qquad (\lambda\in\mathbb{R})\,.
\end{align}
The first and the third solutions are familiar solutions and the last one is the solution of the generalized supergravity equations of motion known in \cite{1903.12175}. 
The second one is an interesting AdS$_3$ solution with a traceful constant $Q$-flux, which is non-Riemannian. 
This seems to be a new solution, but as we explained around Eq.~\eqref{eq:flat-dilaton}, a $B$-shift makes this solution to a flat solution with a linear dilaton, and this is essentially the same as the third solution. 

The correspondence between the Manin triples and the solutions can be summarized as
\begin{align}
\begin{split}
\begin{alignedat}{2}
 &\text{DD2:}\qquad& \underbrace{(\bm{5.i}|\bm{8}|1)}_{\text{Sol1}} &\cong \underbrace{(\bm{8}|\bm{5.i}|1)}_{\text{Sol2}*}\cong
 \underbrace{(\bm{6_0}|\bm{5.iii}|1)}_{\text{Sol3}*} \cong \underbrace{(\bm{5.iii}|\bm{6_0}|1)}_{\text{Sol3}*} \,,
\\
 &\text{DD7:}\qquad& \underbrace{\text{SL(2) WZW}}_{\text{Sol1}} &\cong \underbrace{(\bm{7_0}|\bm{4}|1)}_{\text{Sol4}} \cong \underbrace{(\bm{4}|\bm{7_0}|1)}_{\text{Sol1}}\cong
 \underbrace{(\bm{4}|\bm{2.iii}|1)}_{\text{Sol1}*} \cong \underbrace{(\bm{2.iii}|\bm{4}|1)}_{\text{Sol4}}
\\
 &&&\cong\underbrace{(\bm{6_0}|\bm{4.i}|-1)}_{\text{Sol4}} \cong \underbrace{(\bm{4.i}|\bm{6_0}|-1)}_{\text{Sol1}} \,,
\end{alignedat}
\end{split}
\end{align}
where $\text{SL(2) WZW}$ represents the algebra \eqref{eq:SL2-WZW-alg} and $*$ denotes that Abelian $T$-duality was required in order to bring the non-Riemannian backgrounds into the Riemannian frame or to make solutions of the generalized supergravity to the standard solutions. 

In the case of the NW background, we considered four Manin triples,
\begin{align}
 \text{NW} \cong (E_2^c|A_4) \cong (A_4|E_2^c) \cong (G1|G2) \cong (G2|G1)\,,
\end{align}
where $\text{NW}$ denotes the algebra \eqref{eq:NW-alg}. 
We found that $\text{NW}$, $(E_2^c|A_4)$, and $(G1|G2)$ correspond to the same NW background while $(A_4|E_2^c)$ and $(G2|G1)$ correspond to a non-Riemannian background of the form \eqref{eq:nR-NW2}. 
This non-Riemannian background \eqref{eq:nR-NW2} can be regarded as a flat space with a constant $Q$-flux, and is a kind of $T$-fold if we make some periodic identification of the spatial direction. 

The most interesting solution will be the one obtained by the YB deformation based on the modified CYBE. 
In this case, we found a one-parameter family of solutions which contains the NW background and the flat Minkowski space as particular cases. 

We can apply our procedure to other WZW models, such as the WZW model based on the Heisenberg group $H_4$
\begin{align}
 f_{12}{}^2 = 1\,,\quad
 f_{13}{}^3 = -1\,,\quad
 f_{23}{}^4 = -1\,.
\end{align}
In \cite{1506.06233}, this WZW background was realized as a PL symmetric background by considering a Drinfel'd double
\begin{align}
 f_{12}{}^2 = 1\,,\quad
 f_{13}{}^3 = -1\,,\quad
 f_{23}{}^4 = -1\,,\quad
 f_2{}^{24} = 1\,.
\label{eq:H4-1}
\end{align}
On the other hand, in our approach, the flux algebra is given by
\begin{align}
 \cF_{12}{}^2 = 1\,,\quad
 \cF_{13}{}^3 = -1\,,\quad
 \cF_{23}{}^4 = -1\,,\quad
 \cF_{123}=1\,.
\label{eq:H4-2}
\end{align}
Since the two eight-dimensional algebras \eqref{eq:H4-1} and \eqref{eq:H4-2} are inequivalent, our procedure will provide another way to construct the $H_4$ WZW background based on a new Drinfel'd algebra, and will give a new family of dual geometries. 
Of course, our procedure is not applicable to all of the WZW model. 
For example, if we consider the $\SU(2)$ WZW model, we may start with the algebra
\begin{align}
 \cF_{12}{}^3 = 1\,,\quad
 \cF_{23}{}^1 = 1\,,\quad
 \cF_{31}{}^2 = 1\,,\quad
 \cF_{123} = 1\,.
\end{align}
However, it seems to be impossible to map this algebra into any Lie algebra of a Drinfel'd double through an $\OO(3,3)$ transformation. 
In such cases, our procedure does not work (see \cite{2004.12858} for some discussion on the PL $T$-duality for $\SU(2)$ WZW model). 

In our examples, we found that many Manin triples correspond to a single background. 
As a solution generating technique, this may not seem like a desirable situation, but one can take advantage of this situation \cite{hep-th:9609112}. 
For example, one may exploit the self-duality under PL $T$-plurality in order to search for a new D-brane configuration in a WZW background. 
An embedding of a D-brane can be characterized by the boundary condition of the open string, and the boundary condition on the endpoints of the open string can be characterized by the gluing matrix $R^m{}_n$ which relates the left-/right-moving derivatives on the worldsheet
\begin{align}
 \partial_- x^m = R^m{}_n\,\partial_+ x^n\,.
\end{align}
The gluing matrix for the NW model was studied in \cite{hep-th:9805006} and a similar analysis was done for the AdS$_3$ with $H$-flux in \cite{hep-th:9901122}, and several D-brane configurations were found in these WZW backgrounds. 
Subsequently, the transformation rule of the gluing matrix transforms under the PL $T$-duality was found in \cite{hep-th:0606024} and its extension to the PL $T$-plurality was found in \cite{0706.0820}. 
As we found in this paper, the WZW backgrounds are self-dual under several PL $T$-pluralities, and a naive expectation is that we can find new D-brane configurations by mapping the known gluing matrices through the PL $T$-plurality transformations. 
Moreover, since the AdS$_3$ background is related to the flat space with the linear dilaton, it is also interesting to map the D-brane configurations in these spaces onto each other. 
In addition, our PL $T$-plurality produced many non-Rimannian backgrounds, but D-branes in these backgrounds have not been studied. 
Then it will be interesting to study various D-brane configurations in non-Riemannian backgrounds using the procedure of the PL $T$-plurality studied in \cite{0706.0820}. 

In this paper, we restricted ourselves to the PL $T$-plurality, but the same idea can also be applied to the non-Abelian $U$-duality \cite{1911.06320,1911.07833,2006.12452,2007.08510,2009.04454,2012.13263}. 
For example, in the $E_{n(n)}$ exceptional field theory (EFT) with $n\leq 4$, we may consider a solution of 11D supergravity where the internal parts of the supergravity fields are given by
\begin{align}
 g_{ij} = r_i^a\,r_j^b\,\hat{g}_{ab}\,,\qquad 
 F_4 = \tfrac{1}{4!}\,F_{abcd}\,r^a\wedge\cdots\wedge r^d\qquad (F_4\equiv \rmd C_3)\,,
\end{align}
where $F_{abcd}$ is constant. 
In this case, we can construct the (weightful) generalized metric as $\cM_{IJ} = E_I{}^A\,E_J{}^B\,\hat{\cM}_{AB}$ by using the generalized frame fields
\begin{align}
 E_A{}^I = \begin{pmatrix} E_a{}^I \\ \frac{E^{a_1a_2I}}{\sqrt{2!}} \end{pmatrix} = \begin{pmatrix} e_a^i & -\frac{e_a^k\,C_{ki_1i_2}}{\sqrt{2!}} \\ 0 & r^{a_1}_{[i_1}\,r^{a_2}_{i_2]} \end{pmatrix},
\end{align}
and a constant matrix $\hat{\cM}_{AB}\in \SL(5)$\,. 
We can easily check that this set of the generalized frame fields satisfies the algebra
\begin{align}
 [E_A,\,E_B]_E = - X_{AB}{}^C\,E_C \,,
\label{eq:EDA}
\end{align}
where $[\cdot,\cdot]_E$ is the generalized Lie derivative in EFT and the structure constants are given by
\begin{align}
\begin{split}
 X_{ab}{}^c &= f_{ab}{}^c\,,\quad
 X_{abc_1c_2} = F_{abc_1c_2}\,,\quad
 X_{a}{}^{b_1b_2}{}_{c_1c_2} = 4\,f_{ad}{}^{[b_1}\,\delta^{b_2]d}_{c_1c_2} \,,
\\
 X^{a_1a_2}{}_{b}{}_{c_1c_2} &= 6\,f_{[c_1c_2}{}^{[a_1}\,\delta^{a_2]}_{b]} \,,\quad
 X_{a}{}^{b_1b_2c} =X^{a_1a_2}{}_{b}{}^c = X^{a_1a_2b_1b_2C}=0\,.
\end{split}
\label{eq:EDA}
\end{align}
Under an $\SL(5)$ rotation
\begin{align}
 \hat{\cM}_{AB}\to \hat{\cM}'_{AB} \equiv C_A{}^C\,C_B{}^D\,\hat{\cH}_{CD}\,,\quad
 X_{AB}{}^C\to X'_{AB}{}^C \equiv C_A{}^D\,C_B{}^E\,(C^{-1})_F{}^C\,X_{DE}{}^F\,,
\end{align}
the components $X_{abcd}$ may vanish and the new algebra can be regarded as an exceptional Drinfel'd algebra \cite{1911.06320,1911.07833}. 
In that case, we can construct the new generalized frame fields $E'_A{}^I$ that satisfy the algebra \eqref{eq:EDA} for the structure constants $X'_{AB}{}^C$\,. 
Then we obtain the dual solution $\cM'_{IJ}=E'_I{}^A\,E'_J{}^B\,\hat{\cM}'_{AB}$\,, similar to the PL $T$-plurality discussed in this paper. 

\subsection*{Acknowledgments}

This work is supported by JSPS Grant-in-Aids for Scientific Research (C) 18K13540 and (B) 18H01214.

\end{document}